\documentclass{aa}
\usepackage{xcolor}
\usepackage{graphicx} 
\usepackage{txfonts}
\usepackage{amsmath}

\usepackage{natbib,twoopt}
\usepackage{hyperref}
\usepackage{color}
\usepackage{enumitem}
\usepackage{float}
\usepackage{afterpage}

\usepackage{floatpag}

\usepackage[normalem]{ulem}

\usepackage{placeins}
\usepackage{longtable}
\usepackage{lscape}
\usepackage{lineno}

\usepackage{enumitem}

\usepackage{threeparttable}

\usepackage{placeins}
\usepackage{array}
\usepackage{subcaption}
\usepackage{dcolumn}
\usepackage{booktabs}

\usepackage{siunitx}

\bibpunct{(}{)}{;}{a}{}{,}             

\hypersetup{
  colorlinks=true, 
  urlcolor=blue, 
  linkcolor=blue, 
  citecolor=blue 
}

\newcounter{Rco}
\newcommand{\Ionst}[1]{\setcounter{Rco}{#1}\Roman{Rco}}
\newcommand{\Ion}[2]{\mbox{#1\,{\scriptsize\Ionst{#2}}}}

\newcommand{\teff}{\textit{T}\textsubscript{eff}\xspace}
\newcommand{\logg}{\mbox{log \textit{g}}\xspace}

\newcommand{\teffsol}{\textit{T}\textsubscript{eff}\ensuremath{_{,\odot}}}

\newcommand{\Lsol}{\ensuremath{L_\odot}}
\newcommand{\Msol}{\ensuremath{M_\odot}}
\newcommand{\Rsol}{\ensuremath{R_\odot}}

\newcommand{\Mkiel}{\textit{M}\textsubscript{Kiel}\xspace}

\newcommand{\Mgrav}{\textit{M}\textsubscript{grav}\xspace}

\newcommand{\rad}{\textit{R}\xspace}

\floatpagestyle{empty}

\begin{document}

   \title{Spectral evolution of hot hybrid white dwarfs \\ II. Photometry}

   \author{Semih Filiz\inst{\ref{inst1},\ref{inst2}}
          \and
          Nicole Reindl\inst{\ref{inst1}}
          \and
          David Jones\inst{\ref{inst3},\ref{inst4}}
          \and
          Paulina Sowicka\inst{\ref{inst3},\ref{inst4}}
          \and
          Matti Dorsch\inst{\ref{inst5}}
          \and \\
          Thomas Rauch\inst{\ref{inst2}}
          \and
          Klaus Werner\inst{\ref{inst2}}}
   \institute{Landessternwarte Heidelberg, Zentrum f\"ur Astronomie, Ruprecht-Karls-Universit\"at, K\"onigstuhl 12, 69117 Heidelberg, Germany \\
            \email{sfiliz@lsw.uni-heidelberg.de}\label{inst1}
            \and   
            Institut f\"ur Astronomie und Astrophysik, Kepler Center for Astro and Particle Physics, Eberhard Karls Universit\"at, Sand 1, 72076 T\"ubingen, Germany \label{inst2}
            \and
            Instituto de Astrof\`isica de Canarias, 38205 La Laguna, Tenerife, Spain\label{inst3}
            \and
            Departamento de Astrof\`isica, Universidad de La Laguna, E-38206 La Laguna, Tenerife, Spain\label{inst4}
            \and
            Institut für Physik und Astronomie, Universit\"at Potsdam, Haus 28, Karl-Liebknecht-Str. 24/25, 14476 Potsdam, Germany\label{inst5}
            }
             
   \date{Received 2 September, 2025; accepted 29 October, 2025}

 
  \abstract
   {We present a photometric analysis of 19 DA and 13 DAO white dwarfs (WDs) with effective temperatures exceeding 60\,000\,K, building on the spectral analysis reported in the first paper of this two-part study. By examining archival light curves for periodic signals, we identify that four of the 32 objects ($13^{+8}_{-4}$\%) exhibit photometric variability. Spectral energy distribution (SED) fitting allowed us to derive radii, luminosities, and gravity masses, as well as to characterise the infrared excesses observed in six sources. A notable discovery is the identification of a 1.87\,d period in the ZTF light curves of WD\,1342+443 and weak emission lines in the optical spectra of this star, which strongly indicate an irradiation effect system. Our SED fit indicates the presence of cool dust, which must be located farther from the star, and that any companion with a spectral type earlier than L2.0 would appear in the SED. This leads us to speculate that WD\,1342+443 might have an irradiated, sub-stellar companion. We also highlight that we uncovered, for the first time, a 4.23\,d photometric period in the well-known, close DA+dM binary WD\,0232+035, based on TESS data. We find that the phase and amplitude of the light curve variations are consistent with expectations from an irradiation effect. Intriguingly, we detected an additional, mysterious period at 1.39 days, which is approximately one-third of the orbital period. Moreover, we revisited the longstanding discrepancy between Kiel and gravity masses for the hottest WDs. To address this, we explored fully metal line blanketed model atmospheres as a potential solution, contrasting them with the results from pure-hydrogen and hydrogen-helium models. Our results show that including metal opacities does not resolve the discrepancy -- in fact, it slightly deteriorates the agreement. Finally, we reaffirm the previously observed correlation between helium abundance and luminosity.} 



    \maketitle
\nolinenumbers

\section{Introduction}
\label{sec:Intro}
Large photometric surveys and missions are generally devised for specific research interests in astronomy, such as exploring transient phenomena \citep{2009PASP..121.1395L}, monitoring asteroids and comets \citep{2018PASP..130f4505T}, discovering exoplanets \citep{2015JATIS...1a4003R}, or investigating stellar populations in our Galaxy \citep{2018A&A...619A...4A}. However, the obtained data are useful for purposes other than their intended science objectives, from which white dwarf (WD) studies also benefit substantially. For instance, photometry of a large sample of WDs can be used as a benchmark to validate theories of their internal physical processes \citep{2024Natur.627..286B}. 

On the one hand, time-series observations uncover important aspects of stars, e.g., stellar activity and multiplicity, pulsations, magnetic fields, and rotation. Excluding GW Vir pulsators with inherently hydrogen-deficient atmospheres \citep{2006PASP..118..183W}, close binarity is perhaps the first suspected mechanism that leads to photometric variability in hot WDs \citep{2024ApJ...967..166S,2024ApJ...974..314O}. In particular, so-called irradiation effect systems, in which the phase-varying projection of the cooler companion leads to periodic variations in brightness, are well documented \citep{2006MNRAS.365..287B,2007A&A...469..297A,2010MNRAS.402.2591P}. In these systems, the rotational period of the cool companion is expected to be synchronised to the orbital period \citep{2011A&A...536A..43N}. In recent years, the first indications of bright spots on hot WDs have been reported \citep{2019MNRAS.482L..93R, 2021A&A...647A.184R, 2023A&A...677A..29R}. The partly unusual light curve shapes of these objects are thought to be caused by bright spots on the surfaces of these stars, which in turn are believed to form through interactions of metals with weak magnetic fields \citep{2017ApJ...835..277H}. Furthermore, it is speculated that the onset of diffusion could be an additional decisive factor in the development of spots on hot WDs \citep{2023A&A...677A..29R}.

On the other hand, photometry acquired in multiple bands enables us to construct spectral energy distributions (SEDs). By fitting synthetic spectra to the observed photometry, we can derive the angular diameter of the source. This, in combination with high-precision parallaxes provided by the \emph{Gaia} space mission \citep{2016A&A...595A...2G,2018A&A...616A...1G,2023A&A...674A...1G}, yields stellar radii and, in combination with previously derived surface gravities (\logg), also masses independently of theoretical evolutionary calculations. This method is widely used to derive fundamental parameters for large samples of both H-rich and H-deficient WDs \citep{2019ApJ...876...67B,2019ApJ...871..169G,2019MNRAS.482.5222T}. It has been used to construct a (semi)empirical initial$-$final mass relation \citep{2021AJ....162..162B,2024MNRAS.527.3602C} and to test theoretical predictions of the mass$-$radius relationship \citep{2017ApJ...848...11B,2017MNRAS.465.2849T,2025A&A...695A.131R}. By fitting the SED, it is also  possible to determine the effective temperature (\teff), which remains a free parameter in the fitting procedure along with the angular diameter \citep{2019MNRAS.482.4570G,2021MNRAS.508.3877G}. However, a larger discrepancy in the atmospheric parameters of hot WDs can be identified between photometry and spectroscopy \citep{2019ApJ...876...67B, 2019MNRAS.482.5222T}. Hot WD masses derived from a combination of photometry and spectroscopy also show disagreement with those inferred from evolutionary tracks \citep{2020MNRAS.492..528L,2023A&A...677A..29R}.

Besides offering a method to derive stellar radii and masses independently of evolutionary models, SED fitting also allows us to search for an infrared (IR) excess. Such an excess can indicate the presence of a cooler (sub-)stellar companion, or hot and cold dust discs (\citealt{2011AJ....142...75C, 2013MNRAS.428.2118D, 2024A&A...690A.366R}).

In this paper, we finalise the study reported in \citet[][hereafter Paper I]{2024A&A...691A.290F}, which focuses on the spectral evolution of hot DA and DAO (hybrid) WDs via ultraviolet (UV) and optical spectroscopy. Here, we present a photometric analysis of our sample of WDs that was introduced in Paper I. In Sect.~\ref{sec:LCurves}, we introduce the time-series observations and the analysis method. In Sect.~\ref{sec:SED_fits}, we explain our SED fitting process. We discuss the implications of our results in Sect.~\ref{sec:discussion} and conclude with a brief summary in Sect.~\ref{sec:summary}.

\begin{figure*}[h!]
\centering
\includegraphics[width=17.5cm]{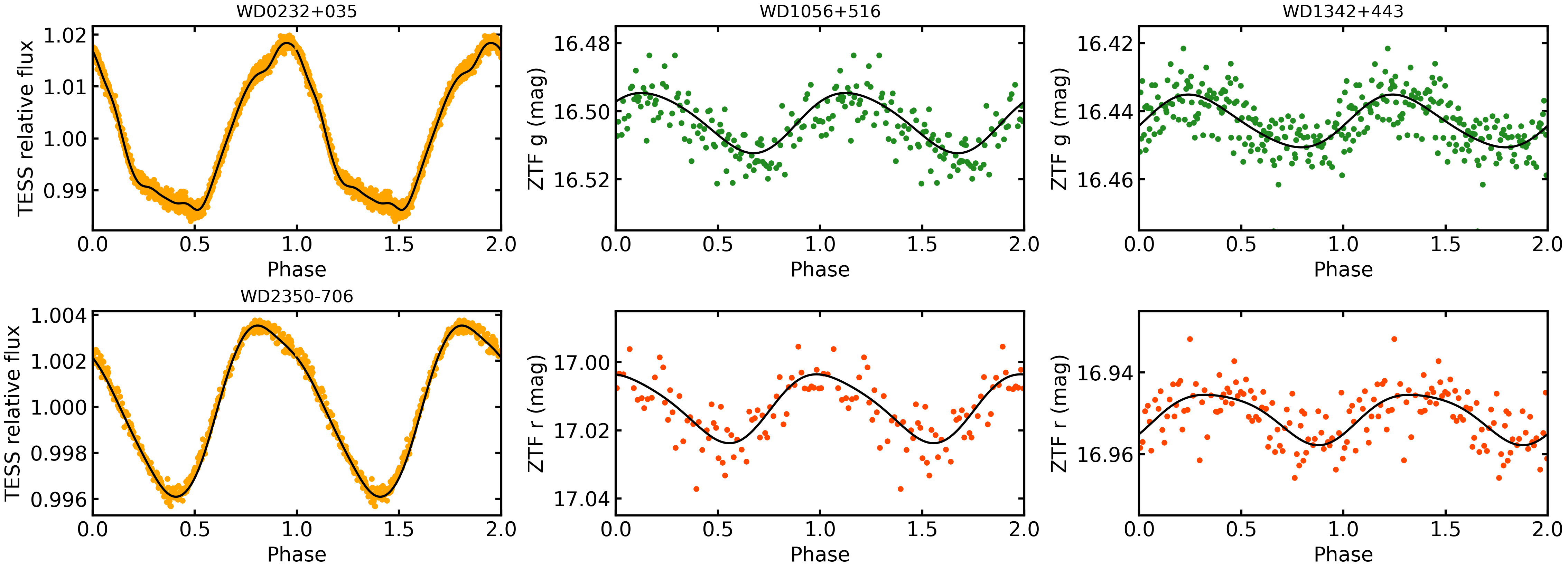}
   \caption{Phase-averaged TESS (orange) and phase-averaged ZTF (green and red) light curves of the variable objects. Peak-to-peak amplitudes were measured by fitting a harmonic series, shown by the black lines. }
     \label{fig:LCall}
\end{figure*}

\section{Light curves}
\label{sec:LCurves} 

\subsection{Datasets}
\label{subsec:datasets}

We collected archival light curves from several surveys, as listed below. All 32 targets from Paper I have at least one archival light curve.

\begin{description}[wide,itemindent=\labelsep]

\item[TESS.]
The Transiting Exoplanet Survey Satellite (TESS) missions comprised individual sectors that were scanned for two orbits ($\approx$\,27.4\,d). The satellite is equipped with a red-optical bandpass covering the range 6\,000 to 10\,000\,\AA. The Science Processing Operations Center (SPOC) pipeline \citep{2016SPIE.9913E..3EJ} provides pre-processed light curves, which we acquired from the Mikulski Archive for Space Telescopes (MAST)\footnote{\url{https://archive.stsci.edu}} along with their corresponding target pixel files (TPFs). We used 2\,min and 20\,s cadence light curves and the pre-search data conditioning simple aperture photometry (PDCSAP) fluxes from which long-term trends have been removed. Two-minute cadence observations are available for all objects, whereas only 21 of them have 20 s cadence data.
\smallskip

\item[ZTF.]
The Zwicky Transient Facility \citep[ZTF;][]{2019PASP..131f8003B,2019PASP..131a8003M} survey is conducted with the Palomar 48-inch telescope at the Palomar Observatory, which scans the northern sky with a wide field of view (FOV; 47\,deg\textsuperscript{2}) at 2 d cadence. We retrieved time-series data with an exposure time of 30\,s in the g and r-band for 21 objects, while i-band observations are available for only ten objects. The data were obtained from the 22nd data release (DR22), spanning March 2018 to June 2024.
\smallskip

\item[CSS.]
The Catalina Sky Survey \citep[CSS;][]{2009ApJ...696..870D} is designed to detect and track near-Earth objects, which employs three 1-m class telescopes located at the Steward Observatory of the University of Arizona. Spanning from 11.5 to 21.5 magnitudes, the second data release of the CSS contains V-band photometry of objects in the range of $-$75° < $\delta$ < 70° and |b| > ~15°. Observations are separated into fields, with each field observed four times every 30 minutes.  We found CSS light curves for 25 objects, but most observations do not have enough data points to detect variability.
\smallskip

\item[ATLAS.]
 The Asteroid Terrestrial-impact Last Alert System \citep[ATLAS,][]{2018PASP..130f4505T} is operated with four automated 0.5 m telescopes (located in Hawaii, South Africa, and Chile) that contain `cyan' (4200$-$6500 \AA) and `orange' (5600$-$8200 \AA) bandpass filters. We found ATLAS light curves only for two of our objects.
\smallskip
\item[\emph{Gaia} multi-epoch photometry.] 
The third data release of the \emph{Gaia} mission \citep{2023A&A...674A...1G} contains multi-epoch broad-band photometry in G, G\textsubscript{BP}, and G\textsubscript{RP} bands, in which variable objects were detected using statistical and machine-learning methods \citep{2023A&A...674A..13E}. Only one object\footnote{Periodicity search for \object{WD\,2353+026} resulted in non-detection.} in our sample has \emph{Gaia} multi-epoch photometry.
\end{description}

\subsection{Analysis}
\label{subsec:LCanalysis}

Our light curve analysis method builds upon the procedure described in \citet{2021A&A...647A.184R,2023A&A...677A..29R,2024A&A...690A.366R}. We used VARTOOLS software \citep{2016A&C....17....1H} and calculated a generalised Lomb-Scargle (LS) periodogram \citep{2009A&A...496..577Z,1992nrca.book.....P} to detect periodic sinusoidal signals. We relied on the false alarm probability (FAP) of the periodic signal to classify variability. If the periodic signal from a particular object showed log(FAP)\,$\leq$\,-4, we classified that object as variable. In cases where the periodogram generated more than one significant period for an object, we removed the strongest periodic signal (including its harmonics and sub-harmonics) from the light curve and recalculated the periodogram. This procedure was repeated until no periodic signal could be found above the set variability threshold (log(FAP)\,$\leq$\,-4). We measured peak-to-peak amplitudes by fitting a harmonic series to each light curve, defined as the difference between the maximum and minimum of the fit. 

The first run of the LS periodogram on the 2\,min cadence TESS data revealed periodic variability for 16 objects. Eight of these objects still showed variability even after the third whitening cycle. In addition, only seven of the variable candidates have a value of \texttt{crowdsap} $\geq$ 0.9, with the median value being 0.74; this parameter in the TPF header estimates the fraction of the collected light in the aperture that comes from the target. In fact, TESS has a large plate scale (21"/pixel), which makes crowding problems unavoidable. Therefore, validating the source of the periodic signal obtained from the light curve is essential.

We inspected the TPFs of the 16 candidate variables mentioned above in all sectors using \texttt{tpfplotter}\footnote{\url{https://github.com/jlillo/tpfplotter}} \citep{2020A&A...635A.128A} to detect possible blending sources or nearby contaminants in the aperture. The TPF plots show the aperture mask used by the pipeline and all sources in the TPF file with a magnitude difference down to $\Delta$m\,=\,6. For only nine objects were no blending sources detected in the aperture of all sectors, whereas in two cases at least one sector did not reveal any contamination in the aperture. However, the absence or scarcity of contamination does not directly imply that the target object is the source of the periodic signal \citep{2024A&A...690A.190A}. Therefore, we used \texttt{TESS$\_$localize}\footnote{\url{https://github.com/Higgins00/TESS-Localize}} \citep{2023AJ....165..141H}, which locates the source of variability on the sky for a given set of frequencies and TESS pixels, as well as the most likely \emph{Gaia} sources within the selected TPF. For detection to be statistically significant, the \texttt{height} parameter should be less than 20\%  \citep[equivalent of 5\,$\sigma$,][]{2023AJ....165..141H}. Consequently, only two objects (\object{WD\,0232+035} and \object{WD\,2350--706}) were attributed to the detected signal. The same procedure was applied to the 20 s cadence data, but the periodogram reproduced the same periods as the 2\,min light curves for only \object{WD\,0232+035} and \object{WD\,0311+480}. However, in all sectors for both datasets, TESS$\_$localize was unable to identify \object{WD\,0311+480} as the source of variability or statistically associate the detected signal with any other \emph{Gaia} source.

We identified periodic signals for only two objects (\object{WD\,1056+516} and \object{WD\,1342+443}) in the ZTF light curves. Both stars show the same periods in the g- and r-bands, but no variability is found in the i-band (229 and 463 data points, respectively). \object{WD\,1056+516} does not show variability in the TESS data. In contrast, we find variability in the TESS data for \object{WD\,1342+443} . However, the period detected in the TESS data does not match that found in the ZTF data and TESS$\_$localize confirms that \object{WD\,1342+443} is not the source of the signal detected in the TESS data. Given that both objects are fainter than G = 16.6 mag and that TESS is designed for objects brighter than 15 mag, this result is not surprising. Fig.~\ref{fig:LCall} shows phase folded light curves of the photometrically variable objects. An overview of the periods and amplitudes derived from the available datasets is shown in Table~\ref{tab:Var_objects}.

\begin{table*}[]
\small
\caption{Names, spectral types, observed bands, false alarm probabilities, periods, and peak--to--peak amplitudes of the variable objects, as well as the temperatures, radii, and surface ratios of the IR-excess sources.}
    \centering
    \setlength{\tabcolsep}{5pt}
    \begin{tabular}{l c l r c c r c c c c}
    \hline\hline\noalign{\smallskip}
    Name    &Spectral & Band & log(FAP) & Period & Amplitude & \textit{T}\textsubscript{eff\,1} & $R_1$ & Surface & \textit{T}\textsubscript{eff\,2} & Surface  \\
            & Type    &      &          &   [d] &  [mag]     & [K] & [\Rsol] & Ratio\textsubscript{1} & [K] & Ratio\textsubscript{2}  \\
    \hline\noalign{\smallskip}
WD\,0232+035   & DA & TESS    &--25834.57 & 4.231915 &  0.032 & $3710^{+70}_{-60}$ & $0.395^{+0.011}_{-0.011}$ & - & - & -\\
               &    &         &--3304.41  & 1.391210 &  0.007 &                  & & & &\\
               &    & ATLAS-c  &  -        & 	-    &  0.029 &                  & & & &\\
               &    & ATLAS-o  & --12.60   & 4.229050 &  0.046 &                 & & & &\\
               &    & CSS-V    &  --9.28   & 4.368181 &  0.160 &                 & & & &\\
\noalign{\smallskip}            
\object{WD\,0851+090}   & DAO& - &  -  &  -   & -        & $2920^{+210}_{-380}$ & $0.260^{+0.040}_{-0.040}$ & - & - & -\\
\noalign{\smallskip}            
\object{WD\,1056+516}   & DA & ZTF-g &--20.15   & 0.113279 &  0.018 & - & - & - & - & -\\
	           &              & ZTF-r &--20.81   & 0.113279 &  0.020 & & & & &\\
	           &              & ZTF-i\,\textsuperscript{a} & -        &    -      & 0.020 & & & & &\\
\noalign{\smallskip}            
\object{WD\,1253+378}   & DAO& -   & -   & - & -    & $3220^{+120}_{-100}$ & $0.337^{+0.022}_{-0.020}$ & - & - & -\\
\noalign{\smallskip}            
\object{WD\,1342+443}   & DA & ZTF-g & --27.60 & 1.873020 &  0.015 & $346^{+17}_{-17}$ & - & $1.4^{+0.5}_{-0.5}$ $\times$ $10^{5}$ & - & - \\
	           &             & ZTF-r & --10.15 & 1.873207 &  0.012 & & & & &\\
	           &             & ZTF-i\,\textsuperscript{a} &  -      &  -       & 0.010 & & & & &\\
\noalign{\smallskip}            
\object{WD\,2218+706}   & DA & - & -   & - & - & $560^{+190}_{-210}$ & - & $1.7^{+11.0}_{-1.1}$ $\times$ $10^{3}$ & $80^{+16}_{-20}$ & $5^{+49}_{-4}$ $\times$ $10^{8}$\\
\noalign{\smallskip}            
\object{WD\,2350$-$706} & DA & TESS & -5306.25  & 1.373420 &  0.008 & $6280^{+50}_{-40}$   & $1.210^{+0.070}_{-0.070}$ & - & - & -\\
\noalign{\smallskip}
\hline
\end{tabular}
\tablefoot{ a. To measure ZTF i-band amplitudes of WD\,1056+516 and WD\,1342+443, we phased the light curves to a significant period found in the respective ZTF g-band.\\
b. Radii of the IR-excess sources were calculated assuming the same distance as the WD.\\
c. Surface ratios are given relative to the WD.}
\label{tab:Var_objects}
\end{table*}

\section{SED fits}
\label{sec:SED_fits}
We used the $\chi$\textsuperscript{2} fitting routine explained in \citet{2018OAst...27...35H} and \citet{2021A&A...650A.102I} to determine the radius, \textit{R}, luminosity, \textit{L}, and gravity mass, \Mgrav, of each sample object. In essence, the procedure converts the model spectra to filter-averaged magnitudes and compares them with the photometric data collected from several surveys and catalogues, including \emph{Gaia} \citep{2021A&A...649A...3R}, the Galaxy Evolution Explorer \citep[GALEX;][]{2005ApJ...619L...1M,2017ApJS..230...24B}, the Panoramic Survey Telescope and Rapid Response System \citep[Pan-STARRS;][]{2020ApJS..251....7F}, Two Micron All Sky Survey \citep[2MASS;][]{2006AJ....131.1163S}, Sloan Digital Sky Survey \citep[SDSS;][]{2015ApJS..219...12A}, the Spitzer Space Telescope \citep{2004ApJS..154...10F,2004ApJS..154...25R}, the Wide-field Infrared Survey Explorer \citep[WISE;][]{2010AJ....140.1868W,2019ApJS..240...30S,2021ApJS..253....8M}, Skymapper \citep{2019PASA...36...33O}, the UK Infra-Red Telescope (UKIRT) Hemisphere Survey \citep[UHS;][]{2018MNRAS.473.5113D} and the Visible and Infrared Survey Telescope Survey \citep[VISTA;][]{2013Msngr.154...35M}. While we fixed the atmospheric parameters (effective temperature, \teff, surface gravity, \logg, and He abundance) to the values determined in Paper I, the angular diameter, $\Theta$, and the colour excess, \textit{E}(44--55), remained free parameters. We adopted the interstellar reddening law of \citet{2019ApJ...886..108F} with an extinction coefficient of \textit{R}\textsubscript{55} = 3.02, which uses \textit{E}(44--55) measured with monochromatic filters at 4400 \AA \xspace and 5500 \AA, rather than the Johnson \textit{B} and \textit{V} filters used in \textit{E(B--V)}. For high \teff, such as our sample objects, \textit{E}(44--55) is indistinguishable from \textit{E(B--V)}.

We relied on \textit{Gaia} parallaxes to calculate the stellar parameters, which were corrected for their zero-point offset according to \citet{2021A&A...649A...4L}. Parallax uncertainties were corrected with the function suggested by \citet{2021MNRAS.506.2269E}. The radii were then calculated as \textit{R} = $\Theta$/ 2$\varpi$, using $\Theta$ derived from the SED fitting and \textit{Gaia} parallaxes ($\varpi$). Using the spectroscopic values (\teff and \logg) from Paper I, we derived the luminosities as \textit{L}/\Lsol = (\rad/\Rsol)\textsuperscript{2}(\teff/\teffsol)\textsuperscript{4} and gravity masses as \Mgrav = \textit{g}$R^{2}$/\textit{G}, where \textit{G} is the gravitational constant.

We followed two different procedures to calculate \rad and \Mgrav using SED fitting. First, we used the model flux of each object from Paper I, computed with the Tübingen Model-Atmosphere package \citep[TMAP\footnote{\url{http://astro.uni-tuebingen.de/~TMAP}};][]{1999JCoAM.109...65W, 2003ASPC..288...31W,2012ascl.soft12015W}, since creating an extensive grid of fully metal line blanketed models is computationally expensive (see Paper I). Then, to test the effects of metals on the SED fits, we used the pure H and H+He model grids of \citet{2023A&A...677A..29R} for the DA and DAO WDs, respectively. Table~\ref{tab:RML} lists the radii, luminosities, gravity masses, and colour excesses derived using the former method. The differences between these results and those obtained using the metal-free models are discussed in further detail in Sect.~\ref{sec:discussion}.

\begin{figure*}[]
     \centering
     \includegraphics[width=17cm]{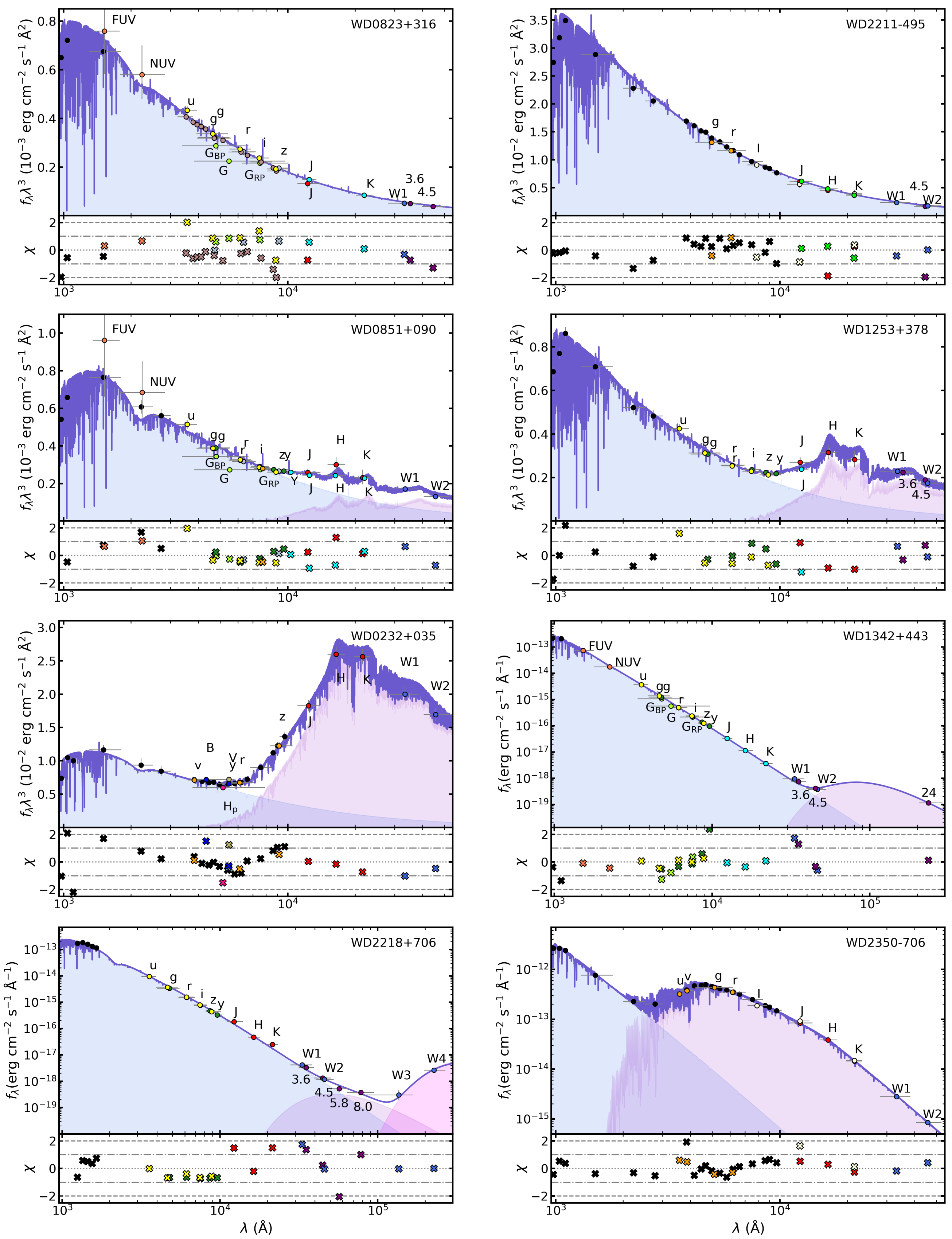}
     \caption{Spectral energy distribution fits to the sample objects (top of each panel) and residuals to the fits (bottom of each panel). The top panels show the SED fits to a DAO (left column) and a DA (right column) WD that do not show an IR excess. The remaining panels (second row: DAO; third and fourth row: DA) depict objects with a near- and/or mid-IR excess modelled with our best fit TMAP models from Paper I and either a best-fit late-type star model from the PHOENIX grid or one or two blackbody component(s). Dots represent observations from different bands, whereas residuals are shown by crosses. Each colour represents a single survey or mission. The dark purple line shows the best-fit (combined) model. Blue and purple areas show the fluxes from the WD and the cool component, respectively. For clarity, the y axis of all panels are shown in the form $f_{\lambda}$\,$\lambda^{3}$, except for WD\,1342+443, WD\,2218+706, and WD\,2350$-$706 where $f_{\lambda}$ is shown on a logarithmic scale.}    
     \label{fig:SED_all}
\end{figure*}

We find that two DAO and four DA WDs in our sample exhibit an IR excess. For WD\,1342+443 and WD\,2218+706, we performed a multi-component fit using the model fluxes obtained in Paper I, together with one or two blackbody components. The other four objects that exhibit an IR excess are already known binaries. For these objects, instead of the blackbody components, we included PHOENIX models \citep{2013A&A...553A...6H} spanning from 500 \AA \xspace to 55\,000 \AA \xspace and temperatures 2300\,K $\leq$ \teff $\leq$ 15\,000\,K. Figure \ref{fig:SED_all} depicts the SED fits for sample objects with IR excess. For comparison, the top row shows one DAO and one DA WD without IR excess. The \teff and \rad of the IR sources are listed in Table~\ref{tab:Var_objects}.

\section{Discussion}
\label{sec:discussion}

\subsection{Variability and IR excess}
\label{sec:remarks}

For all 32 WDs in our sample, at least one archival light curve was available, and four objects were found to be photometrically variable. This implies a variability rate of $13^{+8}_{-4}$\%, which agrees with the variability rate of normal hot, H-rich WDs that do not exhibit the ultra-high excitation (UHE) phenomenon, of $14^{+6}_{-3}$\% as reported by \citet{2021A&A...647A.184R}. In the following paragraphs, we discuss stars that are either (possibly) photometrically variable or show IR excess, considering the origin of the variability and/or the excess.

\begin{figure}[h!]
\resizebox{\hsize}{!}{\includegraphics[trim=0cm 0cm 0cm 0cm, clip]{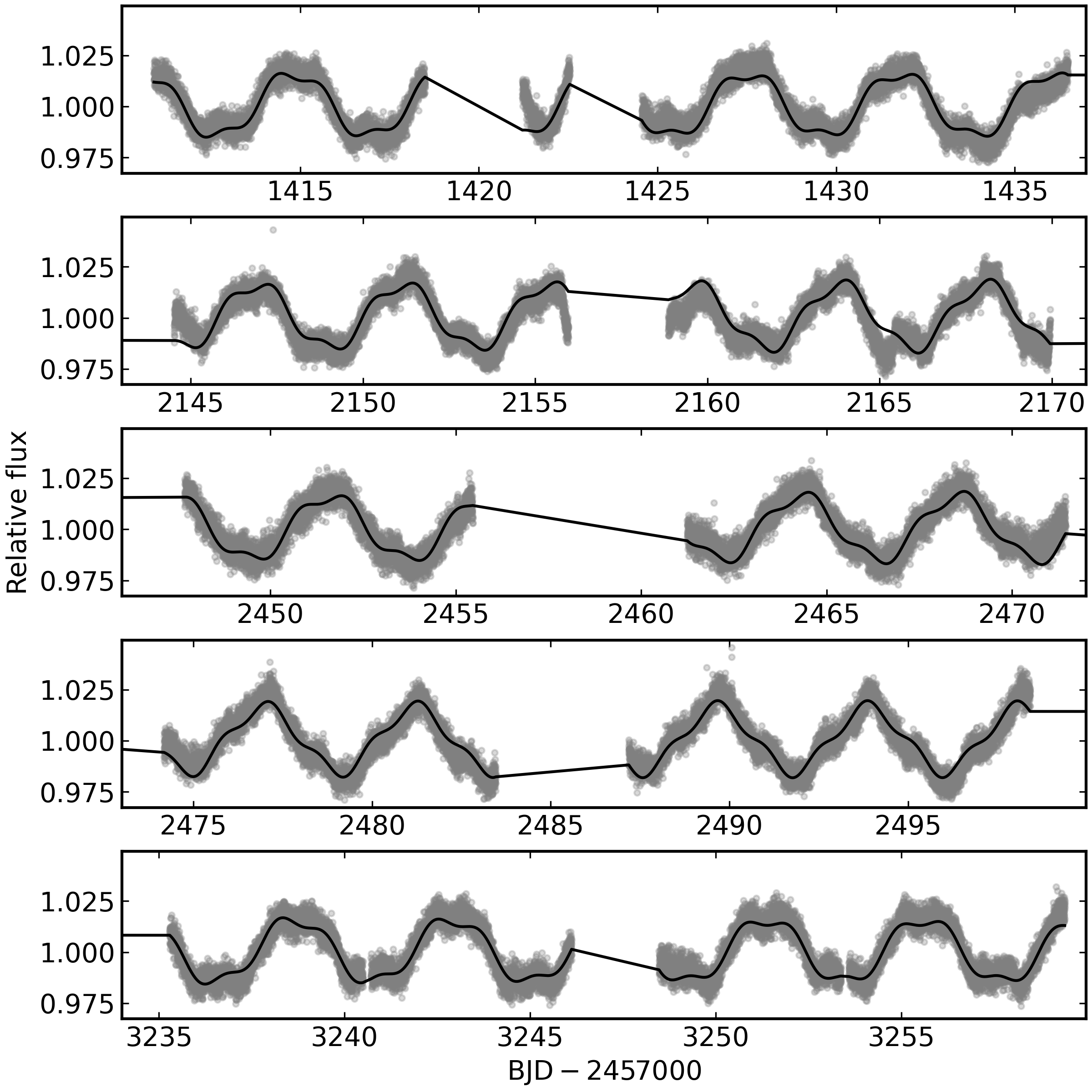}}
  \caption{Time-series TESS photometry (grey) of WD\,0232+035. Each panel represents a TESS sector. A fit of a harmonic series, as a combination of two significant periods (4.23 d and 1.39 d), to the entire TESS dataset is overplotted in black.}
    \label{fig:WD0232_BJD}
\end{figure}

\begin{description}[wide,itemindent=\labelsep]

\item[WD\,0232+035] 

Also known as Feige~24,  WD\,0232+035 is one of the most studied hot DA WDs, with its close binary nature long established \citep{1966VA......8...63G}. Contamination by the irradiated M dwarf is evident in the optical spectra as photospheric absorption (mainly TiO absorption bands) and chromospheric Balmer emission. In previous studies, phase-resolved spectroscopy in the optical and UV enabled the determination of the radial velocity (RV) curves of the M-dwarf and the WD, respectively, yielding an orbital period of 4.23\,d \citep{1991ApJ...372L..37V,1994AJ....108.1881V,2011ApJ...743..138G,2015MNRAS.454.2787G}.\\
We used PHOENIX models to reproduce the IR excess in the SED (Fig.~\ref{fig:SED_all}), resulting in \teff = $3710^{+70}_{-60}$\,K and \rad = 0.395 $\pm$ 0.011\,\Rsol \xspace for the companion. This radius is slightly lower than the average radius of an M dwarf with the corresponding \teff \citep{2020A&A...642A.115C}. \citet{2008ApJ...675.1518K} reported \rad = 0.43 $\pm$ 0.02 \Rsol \xspace for the secondary and classified it as an M2 dwarf.\\ 
We find that the TESS light curve of this object exhibits a 4.23\,d variability, which can be detected in both 2 min and 20 s cadence data. The same period is also present in the o-band ATLAS light curve, but not in the c-band, possibly due to a low number of data points. Therefore, we phased the c-band light curve to the period detected in the o-band. \citet{2018AJ....156..241H} reported a period of 8.45\,d from the ATLAS data and listed the variability as dubious. The LS periodogram on CSS data yielded 0.81\,d as the strongest signal (log(FAP) = -10.75). However, 4.37\,d is also one of the five most significant detected signals (log(FAP) = -9.28), to which we phased the CSS light curve. Moreover, the amplitude measured in the CSS light curves is not consistent with those determined from the TESS and ATLAS data.\\
Although light variation was previously detected with the Hubble Space Telescope (HST) Fine Guidance Sensor 3 \citep{2000AJ....119.2382B}, to the best of our knowledge, the 4.23\,d photometric period is reported here for the first time and is consistent with the system's orbital period inferred from phase-resolved spectroscopy \citep{1978ApJ...223..260T,1991ApJ...372L..37V,1994AJ....108.1881V}.\\ 
The H\,$\alpha$ emission varies with orbital phase, as expected for an irradiation(reflection) effect system; however, the shape of the light curve of WD\,0232+035 (Fig.~\ref{fig:LCall}, lower left panel) does not resemble this at first glance \citep[][see their Fig. 1]{2018A&A...614A..77S}. In contrast, the observed saw-tooth light curve morphology is often indicative of rotational modulation due to spots \citep{2021A&A...647A.184R}.  Interestingly, the light curve morphology does not remain constant throughout the TESS observations (Fig. \ref{fig:WD0232_BJD}), albeit with seemingly periodic repetitions (e.g. data from the start of sector 4, Fig. \ref{fig:WD0232_BJD} upper panel, present a similar light curve morphology as the data from sector 71, Fig. \ref{fig:WD0232_BJD} lower panel). Fitting and subtracting the dominant 4.23\,d periodicity to the light curve using Period04 \citep{2005CoAst.146...53L} reveals a second statistically significant periodicity at 1.39\,d (approximately one third of the orbital period). The beating of these two signals likely causes the variation in light curve morphology.\\
A simulated light curve created with Phoebe2 \citep{2016ApJS..227...29P,2020ApJS..250...34C} suggests that the amplitude and phasing of the photometric variability at the orbital period may be consistent with irradiation. However, the origin of the shorter, lower amplitude 1.39\,d periodicity is less clear. It may be due to spots, but this would imply that the spotted component of the binary is not tidally locked but rather rotating super-synchronously, which is not generally encountered in post-common envelope binaries \citep{2011A&A...536A..43N}. We note that by respectively convolving the flux contributions of the WD and M-dwarf with the TESS spectral response function, we find that both components contribute equally to the observed flux in the TESS band. Intriguingly, the WD-main sequence (MS) binary LAMOST~J172900.17+652952.8 was found to display somewhat similar variability, attributed by \citet[see their Fig.\ 5]{2022ApJ...936...33Z} to strong stellar activity of the MS companion.

\item[WD\,0851+090.] We did not detect any photometric variability for the DAO-type central star of planetary nebula (CSPN) Abell~31, consistent with the findings of \citet{2024A&A...690A.190A} based on their analysis of TESS light curves. However, we find an IR excess for this object (Fig.~\ref{fig:SED_all}), whose origin had previously been attributed to an M4V companion \citep{1999AJ....118..488C,2008PhDT.......109F,2013MNRAS.428.2118D}. We therefore performed a multi-component fit using the PHOENIX grid. We derive \teff = $2920^{+210}_{-380}$ K and \rad = $0.26^{+0.04}_{-0.04}$ \Rsol\xspace for the cool companion, values that are comparable to the average parameters of an M5 dwarf \citep{2020A&A...642A.115C}. The system is not resolved by \emph{Gaia}, but \citet{1999AJ....118..488C} derived a separation of 0.26~arcsec using HST imaging. Using the updated Gaia eDR3 parallax ($\varpi_{Gaia}$ = 1.86~mas), we calculate a separation of 140~AU, which implies that the system is wide but physically bound and most likely did not undergo an interaction phase.

\item[WD\,1056+516.] This DA WD is one of the two objects that show photometric variability in the ZTF dataset. Both g- and r-band light curves show a 0.11 d variability. No period is detected in the i-band (229 data points), likely because of the small amplitude of the light curve variability and/or because the object is fainter in the i-band than in the g- and r-bands. Notably, we find no statistically significant signal in the TESS light curves. However, this object may be too faint for TESS (G = 16.7 mag). Moreover, the low \texttt{crowdsap} value (0.24) indicates that most of the flux detected by TESS originates from nearby sources.\\ 
Since there is no significant difference between the amplitudes between the ZTF g- and r-bands and we did not detect any IR excess in the SED fit, an irradiation effect system seems unlikely. The presence of any hypothetical MS companion can also be excluded, as it would have been revealed by the SED fit. Therefore, the detected variability could be intrinsic. The occurrence of \textit{g}-mode pulsations excited by the $\epsilon$ mechanism is theoretically predicted for hot H-rich post-asymptotic giant branch stars with initial solar metallicity \citep{1988ApJ...334..220K,2014PASJ...66...76M} and for (pre-)WDs for subsolar progenitor metallicity \citep{2017EPJWC.15206012C}. However, the predicted periods ($\sim$\,40 - 200 s) are much shorter than the hour- to day-long periods that we detect for this object. Thus, pulsations can be ruled out as the source of the photometric variability in the present dataset.\\
Because our sample WDs are too hot to have convective atmospheres, a possible explanation for the observed variability is rotational modulation caused by surface inhomogeneities generated by magnetic fields. Aggregated metals near the magnetic poles are predicted to form bright spots in WD atmospheres \citep{2017ApJ...835..277H}, and such features have been detected in several hot WDs with \teff $>$ 30 kK \citep{2017MNRAS.468.1946H}. Weak magnetic fields have also been linked to the UHE phenomenon observed in the optical spectra of hot WDs, which display photometric and spectroscopic variability due to geometrical effects of a wind-fed circumstellar magnetosphere and/or star spots \citep{2019MNRAS.482L..93R,2021A&A...647A.184R}. Although no UHE lines appear present in the optical spectra of WD\,1056+516, the detected periods and measured amplitudes of these two objects are quite similar to those of H-rich UHE variables \citep{2023A&A...677A..29R}. The SDSS spectra show no evidence of Zeeman splitting; however, weaker (below 1\,MG) magnetic field strengths could still lead to the formation of spots \citep{2019MNRAS.482L..93R}. Finally, we note that with a period of 0.11~d, a mass of 0.66\,\Msol, and a radius of 0.011\,\Rsol, the expected rotational velocity is only 5\,km/s, and  rotational broadening would therefore not be detectable in the available spectra.
\smallskip

\item[WD\,1253+378.] 
\citet{2011ApJ...730...67B} reported this DAO WD as a spectroscopic binary with an M5 type companion. Using time-resolved spectroscopy, the same authors could not detect any significant RV variations larger than $20-30$\,km/s. The absence of RV variability is also supported by an earlier study by \cite{2005MNRAS.364.1082G}, who likewise failed to detect RV variations in the UV spectra of this WD. \citet{2011ApJ...730...67B} suggested that the system may either be a longer orbital period binary or a system that is viewed relatively pole-on. \citet{2011AJ....142...75C} did not report any excess in Spitzer 24~$\mu$m observations. The system is not resolved by Gaia, but the renormalised unit weight error (RUWE) from the Gaia DR3 is 1.20. This is slightly higher than the expected value of $\sim$\,1.0 for the single-star astrometric solution \citep{2018A&A...616A...2L,2021A&A...649A...1G}.\\
We did not find photometric variability for WD\,1253+378 in any dataset. This was also confirmed by \citet{2011MNRAS.410..899F} through their analysis of the WASP light curves. \citet{2018AJ....156..241H} reports a period of 2.64~d from ATLAS light curves, but this value is marked as dubious in their catalogue.\\
In our multicomponent SED fit we used PHOENIX models for the M-dwarf companion, finding \teff = $3220^{+120}_{-100}$\,K and \rad = $0.337^{+0.022}_{-0.020}$\,\Rsol. Both the temperature and radius are consistent with those of an M4 dwarf located at the same distance as the WD. Therefore, the system is physically bound but, given the absence of RV and photometric variability, is most likely a wide binary.
\smallskip

\item[WD\,1342+443.] 
We find that this DA WD shows a variability of 1.87~d in the ZTF g- and r-band light curves, which is, to the best of our knowledge, detected for the first time in this study. This period is not confirmed in the TESS data. Instead, we find a variability of 5.42\,d, although TESS$\_$localize did not confirm our WD as the source of this latter period. We stress that the \texttt{crowdsap} value is only 0.26, and given that WD\,1342+443 is relatively faint ($G=16.6$\,mag), the non-detection of the low-amplitude ZTF variability in the TESS data is not surprising.\\
\citet{2011AJ....142...75C} reported an excess in the Spitzer 24~$\mu$m band due to cold dust ($\approx$\,150\,K). We note that an IR excess is already apparent  in the Spitzer IRAC 3.6 and 4.5~$\mu$m band observations \citep{2016MNRAS.459.1415B}, indicating an excess comparable to the W1 and W2 bands. We find that the mid-IR excess in all bands (including the Spitzer 24~$\mu$m band) can be reproduced with a blackbody temperature of 346 $\pm$ 17\,K and a surface ratio of ($1.4^{+0.5}_{-0.4}$) $\times$ $10^{5}$ relative to the WD (Fig.~\ref{fig:SED_all}).\\
The observed mid-IR excess may be caused by cool dust, which is likely the remnant of expelled mass during the asymptotic giant branch phase of the star \citep{2014AJ....147..142C}. Our temperature is slightly higher than that reported by \citet{2011AJ....142...75C}, which can be attributed to their lack of mid-IR observations except for 24~$\mu$m band. This cool dust must be located several AU from the star \citealt{2011AJ....142...75C}), since,  given the high \teff of WD\,1342+443, any dust material close to the star would be sublimated \citep{2007ApJ...662..544V}. A gaseous (debris) disc as the source of the mid-IR excess seems unlikely given the absence of the \ion{Ca}{ii} triplet, which is considered the most important marker of gaseous debris discs \citep{2006Sci...314.1908G}.\\
Although the cool dust scenario explains the mid-IR excess well, it offers no explanation for the observed photometric variability. To search for hints of a close binary nature, we re-investigated the optical spectra of WD\,1342+443. We find that the emission line complex around 4650\,\AA\ is clearly visible in the SDSS and BOSS spectra (Fig.~\ref{fig:BOSS_wd1342}. These lines cannot originate from either the photosphere of the WD (Paper\,I), or from cold dust, whose associated emission lines would not be visible in the optical or NIR range. However, such lines are often seen in irradiation effect systems, where they are produced in the irradiated hemisphere of a cooler companion (e.g. \citealt{2017AJ....153...24H}). Moreover, we identify another emission line at 7726\,\AA\, which may be due to \Ion{Fe}{1} (Fig.~\ref{fig:BOSS_wd1342}). We stress that the signal-to-noise ratio (S/N) of the SDSS sub-spectra is insufficient to check for RV variations in either the weak emission lines or the WD absorption lines.\\
Furthermore, we investigated what type of hypothetical companion could be concealed in the observed SED using the BT-Settl model grid for low-mass stars and brown dwarfs (BDs; \citealt{2011ASPC..448...91A}). Our test indicates that any substellar companion later than L2.0 \citep[$\sim$1800\,K,][]{2020A&A...642A.115C} could be hidden in the SED.\\
Only 0.5\% of WDs are predicted to have a BD companion \citep{2011MNRAS.416.2768S}, and these are generally found around cooler WDs\footnote{Perhaps the only WD that challenges this convention is PG\,1234+482 with \teff $\approx$\,55\,kK. However, the substellar nature of the companion is not confirmed, because the determined spectral type lies on the BD--dM boundary owing to uncertainties in the IR excess \citep{2007MNRAS.382.1804S}.} \citep[\teff $\leq$\,37\,kK,][]{2020ApJ...905..163L}. Several close WD+BD binaries are known in which the substellar companion is irradiated by the WD, and their periods range from $\approx 1\,\mathrm{h} - 1\,\mathrm{d}$ \citep{2018MNRAS.476.1405C,2020MNRAS.499.5318C, 2022ApJ...927L..31R, 2024MNRAS.535..753C}. We speculate that the 1.87~d period detected in WD\,1342+443 may be related to its higher \teff, which could produce an irradiation effect even at longer orbital periods.
\smallskip

\item[WD\,2218+706.] This DA WD was previously considered to be the central star of DeHt~5, later reported to be an HII region ionised by the same WD \citep{2008PhDT.......109F,2013MNRAS.428.2118D,2016MNRAS.455.1459F}. \citet{1996ApJS..107..255T} describe the surrounding environment as a dusty region. Near-IR and mid- IR excesses have been reported for this object \citep{2008PhDT.......109F,2012ApJS..200....3B,2013MNRAS.428.2118D}. Our SED fit also reveals mid-IR excess, which we modelled with two blackbody components. The shorter wavelength component of the mid-IR excess can be reproduced with a blackbody temperature of $560^{+190}_{-210}$\,K and a surface ratio of $1700^{+11000}_{-1100}$ relative to the WD, whereas a temperature of $80^{+16}_{-20}$\,K and a surface ratio of $5^{+49}_{-4}$ $\times$ $10^{8}$ are required to model the W3 and W4 bands. However, a large scatter in the residuals between $\sim$3-6 $\mu$m is noticeable (Fig.\,\ref{fig:SED_all}). These parameters are consistent with the presence of warm and cool dust \citep{2011AJ....142...75C}. We do not detect any photometric variability for this object.
\smallskip

\item[WD\,2350$-$706.] This DA WD is a component of a known non-interacting binary. \citet{1998ApJ...502..763V} reported a constant RV on a short timescale. The system is not resolved by Gaia but was resolved with the HST \citep{2001MNRAS.322..891B}. Using the PHOENIX grid to reproduce the companion in the SED fit, we derive \teff = $6280^{+50}_{-40}$\,K and \textit{R} = $1.21^{+0.07}_{-0.07}$\,\Rsol. These values are consistent with previously reported parameters of an MS F-type star \citep{1994MNRAS.270..499B,1998ApJ...502..763V,2013MNRAS.435.2077H}. Our derived \teff for the companion agrees with the reported values from the sixth data release of the Radial Velocity Experiment (RAVE) survey \citep[\teff = 6267\,K,][]{2020AJ....160...83S} and the third data release of the \emph{Gaia} mission \citep[\teff = 6261\,K,][]{2022yCat.1355....0G,2023A&A...674A...2D}. The 1.38\,d variability we find in the TESS light curve was also reported by \citet{2024AJ....167..189C} and likely corresponds to the rotational period of an F-type star \citep{2021ApJS..255...17S}. Given the 0.574\,arcsec separation reported by \citet{2001MNRAS.322..891B} and the zero-point corrected \emph{Gaia} parallax $\varpi_{Gaia}$ = 6.69 mas, we calculate a separation of 86~AU.
\end{description}

\begin{figure*}[h!]
\sidecaption
\centering
\includegraphics[width=12cm]{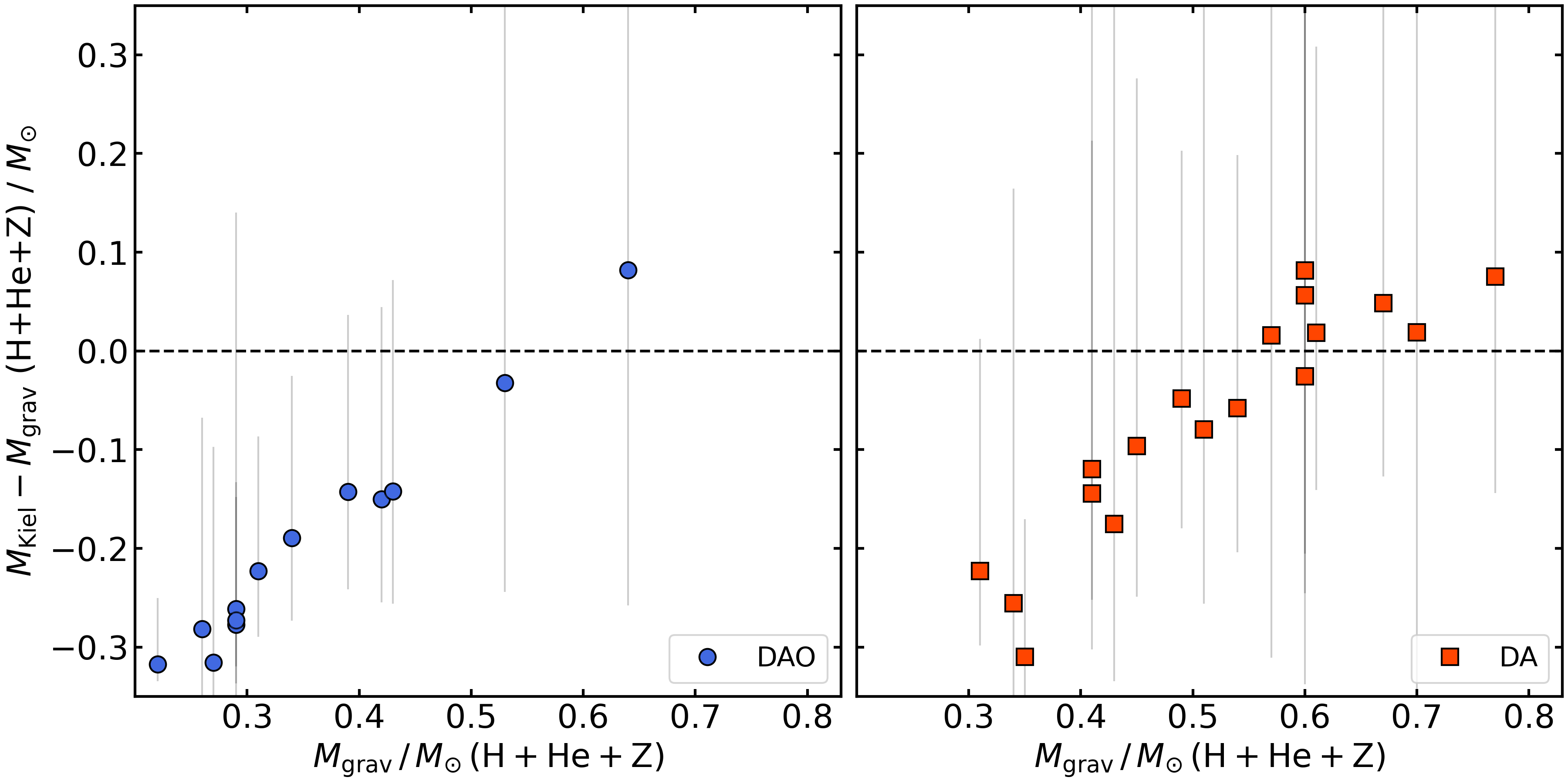}
   \caption{Comparison between gravity and Kiel masses of DAO (blue) and DA (red) WDs. The dashed line represents a 1:1 comparison. Gravity masses were calculated using the radii and surface gravities derived by employing metal line blanketed models in the SED fits and spectral analysis, respectively.}
     \label{fig:MGravZvsMKiel}
\end{figure*}

\begin{figure*}[h!]
\sidecaption
\centering
\includegraphics[width=12cm]{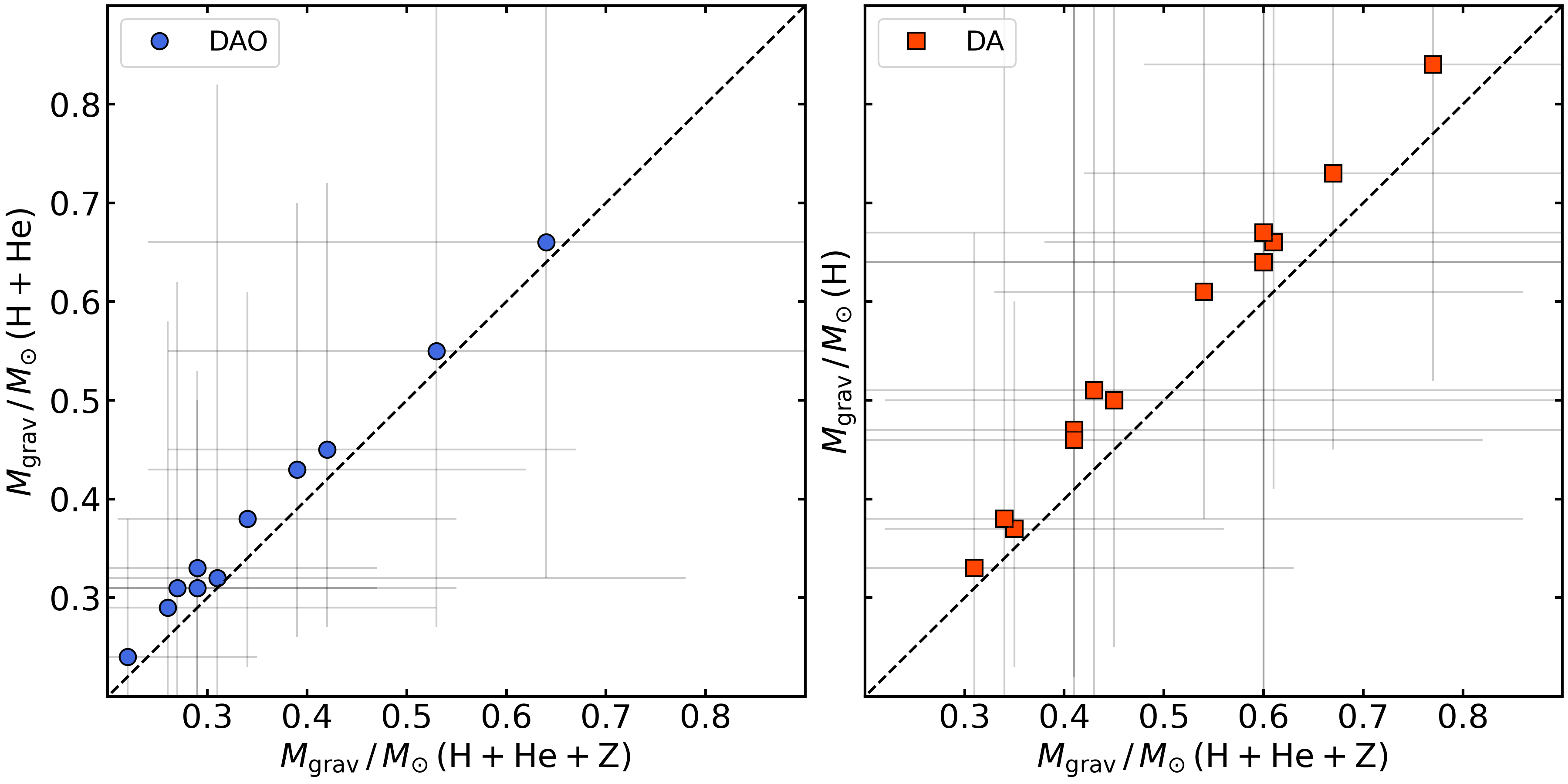}
   \caption{Comparison of gravity masses obtained using metal line blanketed models and pure H and/or H+He models. The dashed line represents a 1:1 comparison.}
     \label{fig:MGravZvsHHe}
\end{figure*}

\subsection{Kiel masses versus gravity masses}
\label{sec:KielvsG}

In Paper I, we used the atmospheric parameters measured from our spectroscopic analysis to interpolate masses from the evolutionary tracks of H-rich WDs by \citet{2010ApJ...717..183R} in the Kiel diagram (\Mkiel). In Sect.~\ref{sec:SED_fits} in this paper, we additionally calculated gravity masses (\Mgrav = \textit{g}$R^{2}$/\textit{G}) using the \logg and \rad obtained from spectral analysis and SED fitting, respectively. Comparing\footnote{WD\,0455$-$282 was excluded from this comparison due to its unusual RUWE value of 8.65, which is not suitable for the single star astrometric solution implemented by \textit{Gaia} \citep{2018A&A...616A...2L,2021A&A...649A...1G}.} the results of these two methods revealed that gravity and Kiel masses agree within 1\,$\sigma$ for 83\% (15/18) of DA and 38\% (5/13) of DAO WDs in our sample .

The conspicuous disparity between Kiel and gravity masses, particularly for the DAO WDs, is evident in Fig.~\ref{fig:MGravZvsMKiel}. This issue manifests systematically as a lower \Mgrav compared to \Mkiel in all cases where a statistical disagreement is identified. It is evident that DAO WDs are more severely affected by this problem than DAs, consistent with the findings of \citet{2023A&A...677A..29R}. We emphasise that unlike \citet{2023A&A...677A..29R} who report only statistical errors from a metal-free $\chi^2$ analysis, our errors also include systematic errors. Despite including systematic errors, the discrepancy between Kiel and gravity masses persists.

We consider Kiel masses to be more reliable than gravity masses. That is because the mean Kiel mass derived in Paper I ($\langle$\textit{M}\textsubscript{Kiel-DAO}$\rangle$ = 0.55 \Msol, $\sigma$ = 0.02 \Msol) is close to the mean Kiel mass of DAs from Paper~I ($\langle$\textit{M}\textsubscript{Kiel-DA}$\rangle$ = 0.59 \Msol, $\sigma$ = 0.05 \Msol) and the mean Kiel masses for DAs reported by \citet{2010ApJ...720..581G}, \citet{2020ApJ...901...93B}, and \citet{2023A&A...677A..29R}. Conversely,  the mean gravity mass for DAOs in our sample (\textit{M}\textsubscript{grav-DAO}$\rangle$ = 0.36 \Msol, $\sigma$ = 0.12 \Msol) is significantly lower than the mean Kiel masses mentioned above, as well as the mean gravity mass for cool hydrogen-rich WDs ($\langle$\textit{M}\textsubscript{grav}$\rangle = 0.61^{+0.14}_{-0.08}$\,\Msol reported by \citet{2025A&A...695A.131R}.

The discrepancy between the mass determination methods based on theoretical tracks and those based on observed parallaxes is a known problem for hot WDs \citep{2020MNRAS.492..528L,2023A&A...677A..29R}. The same issue also appears in other forms. Several studies have reported a mismatch between spectroscopic and parallax distances, with the former generally reported to be larger than the latter \citep{2001A&A...367..973N,2007A&A...470..317R,2012A&A...548A.109Z,2024A&A...690A.366R}. Although in good agreement overall, tests of the theoretical mass--radius relation\footnote{WD\,0232+025 was analysed in all mentioned mass$-$radius studies, whereas WD\,2350$-$706 was only included in the sample of \citet{2018MNRAS.479.1612J}} show that the radii and masses of hot WDs derived from parallax measurements and stellar spectroscopy tend to deviate more from theoretical predictions than those of cooler objects \citep{2017ApJ...848...11B,2017MNRAS.465.2849T,2018MNRAS.479.1612J}. In all of these studies, the larger uncertainties in the atmospheric parameters are considered to be the source of the observed spread, affecting both hot and cool WDs. Revisiting the results of \citet{2017ApJ...848...11B}, \citet{2019ApJ...876...67B} report that, in tests of the theoretical mass$-$radius relationship, the number of objects lying above the 1\,$\sigma$ uncertainty level does not change substantially even after the release of the improved \emph{Gaia} DR2 parallaxes.

Measuring atmospheric parameters of hot WDs is notoriously difficult because the analysis of their optical spectra is heavily hindered by the Balmer line problem (BLP; see Paper I and references therein). \citet{2023A&A...677A..29R} argue that quantitative UV spectral analysis using metal line blanketed nonlocal thermodynamic equilibrium model atmospheres could improve the atmospheric parameters and thereby mitigate the mismatch between the masses determined via different methods. Contrary to these expectations, our study does not support this idea, in which we determined Kiel and gravity masses for the largest sample of the hottest WDs to date using the former method. In fact, including metal opacities seems to slightly worsen the discrepancy between Kiel and gravity masses. This is illustrated in Fig.~\ref{fig:MGravZvsHHe}, where we compare the gravity masses obtained from the SED fits using our fully metal line blanketed models from Paper I, with those obtained from the SED fits using the metal-free H+He grid from \citet{2023A&A...677A..29R}. The SED fits using the metal-free grid predict radii and gravity masses that are on average 5\% and 10\% higher, respectively, than those derived from fully metal line blanketed models. This occurs because models that include metals predict slightly higher fluxes from the far-UV to the near-IR compared to pure H+He models. Given the fixed \teff and \logg, this leads to a lower angular diameter compared to H+He models and, consequently, lower \textit{R} and \Mgrav for a given observed flux and parallax. However, the differences in radii and gravity masses between the two methods are minor.

Neither photometric variability nor the presence of the IR excess appear to be correlated with the mass inconsistency. We also find no correlation between the mismatch with the values we derived for interstellar reddening (Table~\ref{tab:RML}). Conversely, a possible correlation may exist between \teff and the occurrence of mass disagreement. While only 15\% of the objects with \teff $\leq$ 70\,kK show mass disagreement, this value is 50\% for objects above 70\,kK. Since \logg correlates directly with gravity mass, one might argue that \logg values derived in Paper I are systematically underestimated. To test this argument, we increased \logg of the sample WDs by 0.3\,dex and recalculated the Kiel and gravity masses. However, this approach resulted in unrealistically large gravity masses for 65\% of the objects, with differences of up to $\sim$0.8~\Msol\xspace compared to Kiel masses. Although a tighter constraint on \logg might mitigate the problem, a systematic error in the \logg derivation is unlikely.

In addition to the impact of the results from the quantitative spectral analysis, parallax measurements from the \emph{Gaia} mission might also influence the observed discrepancy. We find that 27\% of objects with a parallax measurement larger than 5 mas show a mismatch between \Mgrav and \Mkiel, while this value is 40\% for objects with parallaxes smaller than 5 mas. However, this may simply reflect the fact that the closest objects also tend to be cooler. As parallax increases, it is also unlikely that the mismatch is solely due to the zero-point correction \citep{2023A&A...677A..29R}. We cannot identify a specific cause for the observed mass disagreement. Most likely, a combination of systematic uncertainties in the observations and in the atmospheric parameters is responsible for the detected inconsistency.

\subsection{Evolution of He abundance in the Hertzsprung–Russell diagram} 
\label{sec:HeHRD}

\begin{figure}[]
\resizebox{\hsize}{!}{\includegraphics[trim=0cm 0cm 0cm 0cm, clip]{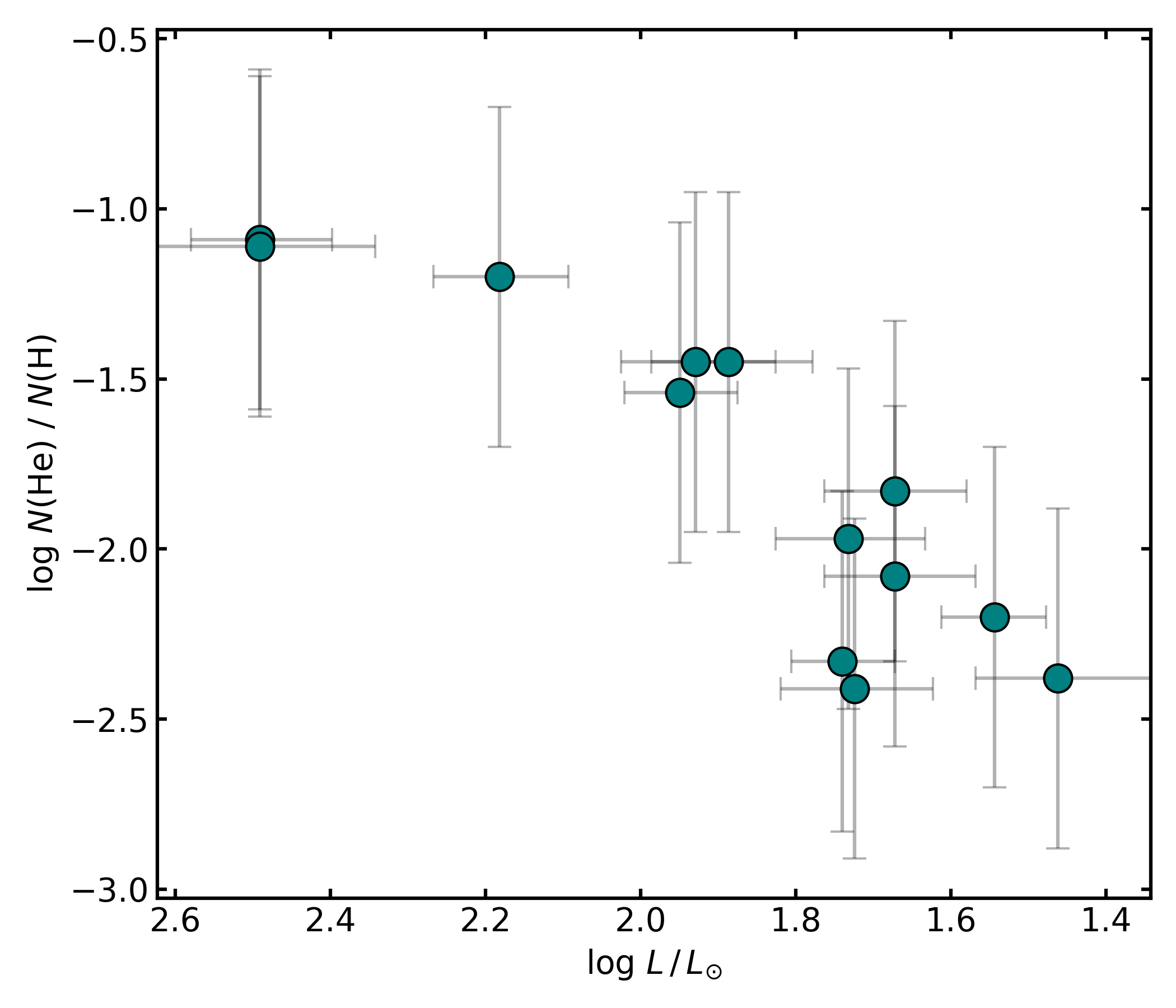}}
  \caption{Helium abundance - luminosity relation of the sample DAO WDs.}
    \label{fig:He_lum}
\end{figure}

We conclude this section by discussing the measured luminosities of our sample WDs. \citet{1999A&A...350..101N} reported a decreasing He abundance with decreasing luminosity in their DAO-type CSPNe sample, which they interpreted as indirect evidence for radiation-driven winds. A similar correlation is evident in our sample, as shown in Fig.~\ref{fig:He_lum}. The decline in He abundance with  luminosity is apparent; at \textit{L} $\approx$ 300~\Lsol, we observe approximately solar He abundance. A sharper decline begins at \textit{L} $\approx$ 100~\Lsol. The characteristic change in He abundance is consistent with previous studies \citep{1999A&A...350..101N,2000A&A...359.1042U}. Some scatter emerges at \textit{L} $\approx$ 65~\Lsol, which is expected and occurs because the theoretically predicted termination of mass loss varies with different masses and initial metallicity \citep{2000A&A...359.1042U}, as was also observationally confirmed in Paper I.

Finally, a clear separation of DAO and DA WDs is evident in the Hertzsprung-Russell diagram (HRD; Fig.~\ref{fig:Kiel_HRD}) consistent with that observed in the Kiel Diagram in Paper I. Our photometric results further reinforce the conclusion of our spectral analysis that all hydrogen-rich WDs are born as DAOs and transform into DA WDs as they cool.

\section{Summary and conclusions}
\label{sec:summary}

We performed a photometric analysis of a sample of 19 DA and 13 DAO WDs with \teff $>60$ kK. All sample objects had at least one archival light curve available from TESS, ZTF, CSS, ATLAS, or \emph{Gaia}. By searching for periodic signals in these datasets, we find that four of the 32 objects in our sample exhibit photometric variability. This corresponds to a variability rate of $13^{+8}_{-4}$\%, which agrees with the variability rate reported for H-rich, non-UHE WDs by \citet{2021A&A...647A.184R}. Furthermore, we analysed the SEDs of all WDs in our sample and identified an IR excess in six of 32 sample objects.\\ 

For the known close DA+dM binary system WD\,0232+035, we report, for the first time, a photometric period of 4.23\,d based on the TESS light curve. We find that the phase and amplitude of the 4.23\,d photometric variability are consistent with expectations for irradiation of the companion. Moreover, we detect an additional lower-amplitude photometric variability with a period of 1.39\,d which is roughly one third of the orbital period. The beating of these two signals produces an unusual light curve shape. The origin of the 1.39\,d period remains unclear. We find that WD\,1056+516 exhibits a 0.11\,d variability in the ZTF g- and r-bands, with similar amplitudes in both bands. We speculate that this variability could be caused by a spot on the surface of the WD, due to the lack of an IR excess. One of the most notable discoveries is the photometric variability of WD\,1342+443, which exhibits similar, low-amplitude variations in the ZTF g- and r-bands with a period of 1.87\,d. We have detected, for the first time, weak emission lines in the optical spectra of this object, which strongly suggests that it is an irradiation effect system. Our SED fit suggests that the mid-IR excess arises from cool dust ($\approx\,350$\,K) that must be located farther from the star. We also demonstrate that any companion with a spectral type earlier than L2.0 would be detectable in the SED, prompting speculation that WD\,1342+443 may host an irradiated sub-stellar companion. Future IR and time-resolved spectroscopic observations are required to better characterise the system. For WD\,2218+760 we show that the IR excess can be explained by warm ($\approx\,560$\,K) and cold ($\approx\,80$\,K) dust. Finally, we suggest that the 1.37\,d period detected in WD\,2350$-$706 likely corresponds to the rotational period of its wide F-type companion.\\

The SED fitting also yielded radii, luminosities and, in combination with \logg from spectroscopy, gravity masses for the WDs. Consequently, the separation of DA and DAO WDs in the \teff$-$\logg plane seen in Paper I is also reproduced in the HRD. In addition, we reinvestigated a longstanding issue in the hottest WDs: the disagreement of Kiel and gravity masses. Specifically, for the first time in a large sample, we assessed if the inclusion of metal opacities in both spectroscopic analysis and SED fitting mitigates this problem. We find that, despite sophisticated metal line blanketed model atmosphere analysis, Kiel and gravity masses agree only in 65\% of cases. In particular, for DAO WDs, the gravity masses appear unrealistically low, and, unexpectedly, including metal opacities mildly exacerbates the problem. Although we cannot resolve this problem, we note that this discrepancy worsens with increasing effective temperatures and possibly decreasing parallaxes. We strongly encourage further investigations into this issue.

\begin{acknowledgements}
We thank the referee for a constructive report that helped to improve the paper. We also thank Veronika Schaffenroth, and JJ Hermes for the helpful discussions. S.F. is supported by the Deutsches Zentrum für Luft- und Raumfahrt (DLR) through grant 50 OR 2315. N.R. is supported by the Deutsche Forschungsgemeinschaft (DFG) through grant RE3915/2-1. D.J. acknowledges support from the Agencia Estatal de Investigaci\'on del Ministerio de Ciencia, Innovaci\'on y Universidades (MCIU/AEI) under grant ``Nebulosas planetarias como clave para comprender la evoluci\'on de estrellas binarias'' and the European Regional Development Fund (ERDF) with reference PID-2022-136653NA-I00 (DOI:10.13039/501100011033). DJ also acknowledges support from the Agencia Estatal de Investigaci\'on del Ministerio de Ciencia, Innovaci\'on y Universidades (MCIU/AEI) under grant ``Revolucionando el conocimiento de la evoluci\'on de estrellas poco masivas'' and the the European Union NextGenerationEU/PRTR with reference CNS2023-143910 (DOI:10.13039/501100011033). P.S. acknowledges support from the Agencia Estatal de Investigación del Ministerio de Ciencia, Innovación y Universidades (MCIU/AEI) under grant “Revolucionando el conocimiento de la evolución de estrellas poco masivas” and the European Union NextGenerationEU/PRTR with reference CNS2023-143910 (DOI:10.13039/501100011033). M.D. was supported by the Deutsches Zentrum für Luft- und Raumfahrt (DLR) through grant 50-OR-2304. The TMAD (\url{http://astro.uni-tuebingen.de/~TMAD}) and TIRO tool (\url{http://astro.uni-tuebingen.de/~TIRO}) used for this paper was constructed as part of the activities of the German Astrophysical Virtual Observatory. This work has made use of data from the European Space Agency (ESA) mission Gaia (\url{https://www.cosmos.esa.int/gaia}), processed by the Gaia Data Processing and Analysis Consortium (DPAC, \url{https://www.cosmos.esa.int/web/gaia/dpac/consortium}). Based on observations obtained with the Samuel Oschin 48-inch Telescope at the Palomar Observatory as part of the Zwicky Transient Facility project. Supported by the National Science Foundation under Grants No. AST-1440341 and AST-2034437 and a collaboration including current partners Caltech, IPAC, the Oskar Klein Center at Stockholm University, the University of Maryland, University of California, Berkeley , the University of Wisconsin at Milwaukee, University of Warwick, Ruhr University, Cornell University, Northwestern University and Drexel University. Operations are conducted by COO, IPAC, and UW. This work includes data from the Asteroid Terrestrial-impact Last Alert System (ATLAS) project. ATLAS is primarily funded to search for near earth asteroids through NASA grants NN12AR55G, 80NSSC18K0284, and 80NSSC18K1575; byproducts of the NEO search include images and catalogs from the survey area. The ATLAS science products have been made possible through the contributions of the University of Hawaii Institute for Astronomy, the Queen’s University Belfast, the Space Telescope Science Institute, and the South African Astronomical Observatory. The CSS survey is funded by the National Aeronautics and Space Administration under Grant No. NNG05GF22G issued through the Science Mission Directorate Near-Earth Objects Observations Program. The CRTS survey is supported by the U.S. National Science Foundation under grants AST0909182 and AST-1313422. The UHS is a partnership between the UK STFC, The University of Hawaii, The University of Arizona, Lockheed Martin and NASA. This paper includes data collected with the TESS mission, obtained from the MAST data archive at the Space Telescope Science Institute (STScI). Funding for the TESS mission is provided by the NASA Explorer Program. STScI is operated by the Association of Universities for Research in Astronomy, Inc., under NASA contract NAS 5–26555. This work made use of tpfplotter by J. Lillo-Box (publicly available in \url{www.github.com/jlillo/tpfplotter}), which also made use of the python packages astropy, lightkurve, matplotlib and numpy. This research has made use of NASA’s Astrophysics Data System and the SIMBAD database, operated at CDS, Strasbourg, France. This research has made use of the VizieR catalogue access tool, CDS, Strasbourg, France. This research made use of TOPCAT, an interactive graphical viewer and editor for tabular data \citep{2005ASPC..347...29T}. 
\end{acknowledgements}

\bibliographystyle{aa} 
\bibliography{bibfile} 

@ARTICLE{2013MNRAS.428.2118D,
       author = {{De Marco}, Orsola and {Passy}, Jean-Claude and {Frew}, D.~J. and {Moe}, Maxwell and {Jacoby}, G.~H.},
        title = "{The binary fraction of planetary nebula central stars - I. A high-precision, I-band excess search}",
      journal = {\mnras},
         year = 2013,
        month = jan,
       volume = {428},
       number = {3},
        pages = {2118-2140},
          doi = {10.1093/mnras/sts180},
archivePrefix = {arXiv},
       eprint = {1210.2841},
 primaryClass = {astro-ph.SR},
       adsurl = {https://ui.adsabs.harvard.edu/abs/2013MNRAS.428.2118D}
}

@ARTICLE{2022ApJ...927L..31R,
       author = {{Rebassa-Mansergas}, Alberto and {Xu}, Siyi and {Raddi}, Roberto and {Pala}, Anna F. and {Solano}, Enrique and {Torres}, Santiago and {Jim{\'e}nez-Esteban}, Francisco and {Cruz}, Patricia},
        title = "{Gaia 0007-1605: An Old Triple System with an Inner Brown Dwarf-White Dwarf Binary and an Outer White Dwarf Companion}",
      journal = {\apjl},
         year = 2022,
        month = mar,
       volume = {927},
       number = {2},
          eid = {L31},
        pages = {L31},
          doi = {10.3847/2041-8213/ac5a55},
archivePrefix = {arXiv},
       eprint = {2203.05901},
 primaryClass = {astro-ph.SR},
       adsurl = {https://ui.adsabs.harvard.edu/abs/2022ApJ...927L..31R}
}

@ARTICLE{2017AJ....153...24H,
       author = {{Hillwig}, Todd C. and {Frew}, David J. and {Reindl}, Nicole and {Rotter}, Hannah and {Webb}, Andrew and {Margheim}, Steve},
        title = "{Binary Central Stars of Planetary Nebulae Discovered Through Photometric Variability. V. The Central Stars of HaTr 7 and ESO 330-9}",
      journal = {\aj},
         year = 2017,
        month = jan,
       volume = {153},
       number = {1},
          eid = {24},
        pages = {24},
          doi = {10.3847/1538-3881/153/1/24},
archivePrefix = {arXiv},
       eprint = {1612.01420},
 primaryClass = {astro-ph.SR},
       adsurl = {https://ui.adsabs.harvard.edu/abs/2017AJ....153...24H}
}

@ARTICLE{2019MNRAS.482L..93R,
       author = {{Reindl}, Nicole and {Bainbridge}, M. and {Przybilla}, N. and {Geier}, S. and {Prv{\'a}k}, M. and {Krti{\v{c}}ka}, J. and {{\O}stensen}, R.~H. and {Telting}, J. and {Werner}, K.},
        title = "{Unravelling the baffling mystery of the ultrahot wind phenomenon in white dwarfs}",
      journal = {\mnras},
         year = 2019,
        month = jan,
       volume = {482},
       number = {1},
        pages = {L93-L98},
          doi = {10.1093/mnrasl/sly191},
archivePrefix = {arXiv},
       eprint = {1811.02922},
 primaryClass = {astro-ph.SR},
       adsurl = {https://ui.adsabs.harvard.edu/abs/2019MNRAS.482L..93R}
}

@ARTICLE{2011AJ....142...75C,
       author = {{Chu}, You-Hua and {Su}, Kate Y.~L. and {Bilikova}, Jana and {Gruendl}, Robert A. and {De Marco}, Orsola and {Guerrero}, Martin A. and {Updike}, Adria C. and {Volk}, Kevin and {Rauch}, Thomas},
        title = "{Spitzer 24 {\ensuremath{\mu}}m Survey for Dust Disks around Hot White Dwarfs}",
      journal = {\aj},
         year = 2011,
        month = sep,
       volume = {142},
       number = {3},
          eid = {75},
        pages = {75},
          doi = {10.1088/0004-6256/142/3/75},
archivePrefix = {arXiv},
       eprint = {1101.5137},
 primaryClass = {astro-ph.SR},
       adsurl = {https://ui.adsabs.harvard.edu/abs/2011AJ....142...75C}
}

@INPROCEEDINGS{2005ASPC..347...29T,
       author = {{Taylor}, M.~B.},
        title = "{TOPCAT \& STIL: Starlink Table/VOTable Processing Software}",
    booktitle = {Astronomical Data Analysis Software and Systems XIV},
         year = 2005,
       editor = {{Shopbell}, P. and {Britton}, M. and {Ebert}, R.},
       series = {Astronomical Society of the Pacific Conference Series},
       volume = {347},
        month = dec,
        pages = {29},
       adsurl = {https://ui.adsabs.harvard.edu/abs/2005ASPC..347...29T}
}

@ARTICLE{2005CoAst.146...53L,
       author = {{Lenz}, P. and {Breger}, M.},
        title = "{Period04 User Guide}",
      journal = {Communications in Asteroseismology},
         year = 2005,
        month = jun,
       volume = {146},
        pages = {53-136},
          doi = {10.1553/cia146s53},
       adsurl = {https://ui.adsabs.harvard.edu/abs/2005CoAst.146...53L}
}

@ARTICLE{2020ApJS..250...34C,
       author = {{Conroy}, Kyle E. and {Kochoska}, Angela and {Hey}, Daniel and {Pablo}, Herbert and {Hambleton}, Kelly M. and {Jones}, David and {Giammarco}, Joseph and {Abdul-Masih}, Michael and {Pr{\v{s}}a}, Andrej},
        title = "{Physics of Eclipsing Binaries. V. General Framework for Solving the Inverse Problem}",
      journal = {\apjs},
         year = 2020,
        month = oct,
       volume = {250},
       number = {2},
          eid = {34},
        pages = {34},
          doi = {10.3847/1538-4365/abb4e2},
archivePrefix = {arXiv},
       eprint = {2006.16951},
 primaryClass = {astro-ph.SR},
       adsurl = {https://ui.adsabs.harvard.edu/abs/2020ApJS..250...34C}
}

@ARTICLE{2022ApJ...936...33Z,
       author = {{Zheng}, Ling-Lin and {Gu}, Wei-Min and {Sun}, Mouyuan and {Zhang}, Zhi-Xiang and {Yi}, Tuan and {Wu}, Jianfeng and {Wang}, Junfeng and {Fu}, Jin-Bo and {Qi}, Sen-Yu and {Yang}, Fan and {Wang}, Song and {Wang}, Liang and {Bai}, Zhong-Rui and {Zhang}, Haotong and {Li}, Chun-Qian and {Shi}, Jian-Rong and {Zong}, Weikai and {Bai}, Yu and {Liu}, Jifeng},
        title = "{A White Dwarf-Main-sequence Binary Unveiled by Time-domain Observations from LAMOST and TESS}",
      journal = {\apj},
         year = 2022,
        month = sep,
       volume = {936},
       number = {1},
          eid = {33},
        pages = {33},
          doi = {10.3847/1538-4357/ac853f},
archivePrefix = {arXiv},
       eprint = {2209.13924},
 primaryClass = {astro-ph.SR},
       adsurl = {https://ui.adsabs.harvard.edu/abs/2022ApJ...936...33Z}
}

@ARTICLE{2016ApJS..227...29P,
       author = {{Pr{\v{s}}a}, A. and {Conroy}, K.~E. and {Horvat}, M. and {Pablo}, H. and {Kochoska}, A. and {Bloemen}, S. and {Giammarco}, J. and {Hambleton}, K.~M. and {Degroote}, P.},
        title = "{Physics Of Eclipsing Binaries. II. Toward the Increased Model Fidelity}",
      journal = {\apjs},
         year = 2016,
        month = dec,
       volume = {227},
       number = {2},
          eid = {29},
        pages = {29},
          doi = {10.3847/1538-4365/227/2/29},
archivePrefix = {arXiv},
       eprint = {1609.08135},
 primaryClass = {astro-ph.SR},
       adsurl = {https://ui.adsabs.harvard.edu/abs/2016ApJS..227...29P}
}

@ARTICLE{2010ApJ...720..581G,
       author = {{Gianninas}, A. and {Bergeron}, P. and {Dupuis}, J. and {Ruiz}, M.~T.},
        title = "{Spectroscopic Analysis of Hot, Hydrogen-rich White Dwarfs: The Presence of Metals and the Balmer-line Problem}",
      journal = {\apj},
         year = 2010,
        month = sep,
       volume = {720},
       number = {1},
        pages = {581-602},
          doi = {10.1088/0004-637X/720/1/581},
       adsurl = {https://ui.adsabs.harvard.edu/abs/2010ApJ...720..581G}
}

@ARTICLE{2011ApJ...743..138G,
       author = {{Gianninas}, A. and {Bergeron}, P. and {Ruiz}, M.~T.},
        title = "{A Spectroscopic Survey and Analysis of Bright, Hydrogen-rich White Dwarfs}",
      journal = {\apj},
         year = 2011,
        month = dec,
       volume = {743},
       number = {2},
          eid = {138},
        pages = {138},
          doi = {10.1088/0004-637X/743/2/138},
archivePrefix = {arXiv},
       eprint = {1109.3171},
 primaryClass = {astro-ph.SR},
       adsurl = {https://ui.adsabs.harvard.edu/abs/2011ApJ...743..138G}
}

@ARTICLE{2010ApJ...717..183R,
       author = {{Renedo}, I. and {Althaus}, L.~G. and {Miller Bertolami}, M.~M. and {Romero}, A.~D. and {C{\'o}rsico}, A.~H. and {Rohrmann}, R.~D. and {Garc{\'\i}a-Berro}, E.},
        title = "{New Cooling Sequences for Old White Dwarfs}",
      journal = {\apj},
         year = 2010,
        month = jul,
       volume = {717},
       number = {1},
        pages = {183-195},
          doi = {10.1088/0004-637X/717/1/183},
archivePrefix = {arXiv},
       eprint = {1005.2170},
 primaryClass = {astro-ph.SR},
       adsurl = {https://ui.adsabs.harvard.edu/abs/2010ApJ...717..183R}
}

@ARTICLE{2000A&A...359.1042U,
       author = {{Unglaub}, K. and {Bues}, I.},
        title = "{The chemical evolution of hot white dwarfs in the presence of diffusion and mass loss}",
      journal = {\aap},
         year = 2000,
        month = jul,
       volume = {359},
        pages = {1042-1058},
       adsurl = {https://ui.adsabs.harvard.edu/abs/2000A&A...359.1042U}
}

@ARTICLE{2020ApJ...901...93B,
       author = {{B{\'e}dard}, A. and {Bergeron}, P. and {Brassard}, P. and {Fontaine}, G.},
        title = "{On the Spectral Evolution of Hot White Dwarf Stars. I. A Detailed Model Atmosphere Analysis of Hot White Dwarfs from SDSS DR12}",
      journal = {\apj},
         year = 2020,
        month = oct,
       volume = {901},
       number = {2},
          eid = {93},
        pages = {93},
          doi = {10.3847/1538-4357/abafbe},
archivePrefix = {arXiv},
       eprint = {2008.07469},
 primaryClass = {astro-ph.SR},
       adsurl = {https://ui.adsabs.harvard.edu/abs/2020ApJ...901...93B}
}

@ARTICLE{2016A&C....17....1H,
       author = {{Hartman}, J.~D. and {Bakos}, G. {\'A}.},
        title = "{VARTOOLS: A program for analyzing astronomical time-series data}",
      journal = {Astronomy and Computing},
         year = 2016,
        month = oct,
       volume = {17},
        pages = {1-72},
          doi = {10.1016/j.ascom.2016.05.006},
archivePrefix = {arXiv},
       eprint = {1605.06811},
 primaryClass = {astro-ph.IM},
       adsurl = {https://ui.adsabs.harvard.edu/abs/2016A&C....17....1H}
}

@BOOK{1992nrca.book.....P,
       author = {{Press}, William H. and {Teukolsky}, Saul A. and {Vetterling}, William T. and {Flannery}, Brian P.},
        title = "{Numerical recipes in C. The art of scientific computing}",
         year = 1992,
       adsurl = {https://ui.adsabs.harvard.edu/abs/1992nrca.book.....P}
}

@ARTICLE{2009A&A...496..577Z,
       author = {{Zechmeister}, M. and {K{\"u}rster}, M.},
        title = "{The generalised Lomb-Scargle periodogram. A new formalism for the floating-mean and Keplerian periodograms}",
      journal = {\aap},
         year = 2009,
        month = mar,
       volume = {496},
       number = {2},
        pages = {577-584},
          doi = {10.1051/0004-6361:200811296},
archivePrefix = {arXiv},
       eprint = {0901.2573},
 primaryClass = {astro-ph.IM},
       adsurl = {https://ui.adsabs.harvard.edu/abs/2009A&A...496..577Z}
}

@ARTICLE{2021A&A...647A.184R,
       author = {{Reindl}, Nicole and {Schaffenroth}, Veronika and {Filiz}, Semih and {Geier}, Stephan and {Pelisoli}, Ingrid and {Kepler}, Souza Oliveira},
        title = "{Mysterious, variable, and extremely hot: White dwarfs showing ultra-high excitation lines. I. Photometric variability}",
      journal = {\aap},
         year = 2021,
        month = mar,
       volume = {647},
          eid = {A184},
        pages = {A184},
          doi = {10.1051/0004-6361/202140289},
archivePrefix = {arXiv},
       eprint = {2102.04504},
 primaryClass = {astro-ph.SR},
       adsurl = {https://ui.adsabs.harvard.edu/abs/2021A&A...647A.184R}
}

@ARTICLE{2023A&A...677A..29R,
       author = {{Reindl}, Nicole and {Islami}, Ramazan and {Werner}, Klaus and {Kepler}, S.~O. and {Pritzkuleit}, Max and {Dawson}, Harry and {Dorsch}, Matti and {Istrate}, Alina and {Pelisoli}, Ingrid and {Geier}, Stephan and {Uzundag}, Murat and {Provencal}, Judith and {Justham}, Stephen},
        title = "{The bright blue side of the night sky: Spectroscopic survey of bright and hot (pre-) white dwarfs}",
      journal = {\aap},
         year = 2023,
        month = sep,
       volume = {677},
          eid = {A29},
        pages = {A29},
          doi = {10.1051/0004-6361/202346865},
archivePrefix = {arXiv},
       eprint = {2307.03721},
 primaryClass = {astro-ph.SR},
       adsurl = {https://ui.adsabs.harvard.edu/abs/2023A&A...677A..29R}
}

@ARTICLE{2021A&A...650A.102I,
       author = {{Irrgang}, A. and {Geier}, S. and {Heber}, U. and {Kupfer}, T. and {El-Badry}, K. and {Bloemen}, S.},
        title = "{A proto-helium white dwarf stripped by a substellar companion via common-envelope ejection. Uncovering the true nature of a candidate hypervelocity B-type star}",
      journal = {\aap},
         year = 2021,
        month = jun,
       volume = {650},
          eid = {A102},
        pages = {A102},
          doi = {10.1051/0004-6361/202038757},
archivePrefix = {arXiv},
       eprint = {2007.03350},
 primaryClass = {astro-ph.SR},
       adsurl = {https://ui.adsabs.harvard.edu/abs/2021A&A...650A.102I}
}

@ARTICLE{2018OAst...27...35H,
       author = {{Heber}, Ulrich and {Irrgang}, Andreas and {Schaffenroth}, Johannes},
        title = "{Spectral energy distributions and colours of hot subluminous stars}",
      journal = {Open Astronomy},
         year = 2018,
        month = feb,
       volume = {27},
       number = {1},
        pages = {35-43},
          doi = {10.1515/astro-2018-0008},
archivePrefix = {arXiv},
       eprint = {1712.06546},
 primaryClass = {astro-ph.SR},
       adsurl = {https://ui.adsabs.harvard.edu/abs/2018OAst...27...35H}
}

@ARTICLE{2019ApJ...886..108F,
       author = {{Fitzpatrick}, E.~L. and {Massa}, Derck and {Gordon}, Karl D. and {Bohlin}, Ralph and {Clayton}, Geoffrey C.},
        title = "{An Analysis of the Shapes of Interstellar Extinction Curves. VII. Milky Way Spectrophotometric Optical-through-ultraviolet Extinction and Its R-dependence}",
      journal = {\apj},
         year = 2019,
        month = dec,
       volume = {886},
       number = {2},
          eid = {108},
        pages = {108},
          doi = {10.3847/1538-4357/ab4c3a},
archivePrefix = {arXiv},
       eprint = {1910.08852},
 primaryClass = {astro-ph.GA},
       adsurl = {https://ui.adsabs.harvard.edu/abs/2019ApJ...886..108F}
}

@ARTICLE{2018A&A...619A...4A,
       author = {{Alonso-Garc{\'\i}a}, Javier and {Saito}, Roberto K. and {Hempel}, Maren and {Minniti}, Dante and {Pullen}, Joyce and {Catelan}, M{\'a}rcio and {Ramos}, Rodrigo Contreras and {Cross}, Nicholas J.~G. and {Gonzalez}, Oscar A. and {Lucas}, Philip W. and {Palma}, Tali and {Valenti}, Elena and {Zoccali}, Manuela},
        title = "{Milky Way demographics with the VVV survey. IV. PSF photometry from almost one billion stars in the Galactic bulge and adjacent southern disk}",
      journal = {\aap},
         year = 2018,
        month = oct,
       volume = {619},
          eid = {A4},
        pages = {A4},
          doi = {10.1051/0004-6361/201833432},
archivePrefix = {arXiv},
       eprint = {1808.06139},
 primaryClass = {astro-ph.GA},
       adsurl = {https://ui.adsabs.harvard.edu/abs/2018A&A...619A...4A}
}

@ARTICLE{2009PASP..121.1395L,
       author = {{Law}, Nicholas M. and {Kulkarni}, Shrinivas R. and {Dekany}, Richard G. and {Ofek}, Eran O. and {Quimby}, Robert M. and {Nugent}, Peter E. and {Surace}, Jason and {Grillmair}, Carl C. and {Bloom}, Joshua S. and {Kasliwal}, Mansi M. and {Bildsten}, Lars and {Brown}, Tim and {Cenko}, S. Bradley and {Ciardi}, David and {Croner}, Ernest and {Djorgovski}, S. George and {van Eyken}, Julian and {Filippenko}, Alexei V. and {Fox}, Derek B. and {Gal-Yam}, Avishay and {Hale}, David and {Hamam}, Nouhad and {Helou}, George and {Henning}, John and {Howell}, D. Andrew and {Jacobsen}, Janet and {Laher}, Russ and {Mattingly}, Sean and {McKenna}, Dan and {Pickles}, Andrew and {Poznanski}, Dovi and {Rahmer}, Gustavo and {Rau}, Arne and {Rosing}, Wayne and {Shara}, Michael and {Smith}, Roger and {Starr}, Dan and {Sullivan}, Mark and {Velur}, Viswa and {Walters}, Richard and {Zolkower}, Jeff},
        title = "{The Palomar Transient Factory: System Overview, Performance, and First Results}",
      journal = {\pasp},
         year = 2009,
        month = dec,
       volume = {121},
       number = {886},
        pages = {1395},
          doi = {10.1086/648598},
archivePrefix = {arXiv},
       eprint = {0906.5350},
 primaryClass = {astro-ph.IM},
       adsurl = {https://ui.adsabs.harvard.edu/abs/2009PASP..121.1395L}
}

@ARTICLE{2015JATIS...1a4003R,
       author = {{Ricker}, George R. and {Winn}, Joshua N. and {Vanderspek}, Roland and {Latham}, David W. and {Bakos}, G{\'a}sp{\'a}r {\'A}. and {Bean}, Jacob L. and {Berta-Thompson}, Zachory K. and {Brown}, Timothy M. and {Buchhave}, Lars and {Butler}, Nathaniel R. and {Butler}, R. Paul and {Chaplin}, William J. and {Charbonneau}, David and {Christensen-Dalsgaard}, J{\o}rgen and {Clampin}, Mark and {Deming}, Drake and {Doty}, John and {De Lee}, Nathan and {Dressing}, Courtney and {Dunham}, Edward W. and {Endl}, Michael and {Fressin}, Francois and {Ge}, Jian and {Henning}, Thomas and {Holman}, Matthew J. and {Howard}, Andrew W. and {Ida}, Shigeru and {Jenkins}, Jon M. and {Jernigan}, Garrett and {Johnson}, John Asher and {Kaltenegger}, Lisa and {Kawai}, Nobuyuki and {Kjeldsen}, Hans and {Laughlin}, Gregory and {Levine}, Alan M. and {Lin}, Douglas and {Lissauer}, Jack J. and {MacQueen}, Phillip and {Marcy}, Geoffrey and {McCullough}, Peter R. and {Morton}, Timothy D. and {Narita}, Norio and {Paegert}, Martin and {Palle}, Enric and {Pepe}, Francesco and {Pepper}, Joshua and {Quirrenbach}, Andreas and {Rinehart}, Stephen A. and {Sasselov}, Dimitar and {Sato}, Bun'ei and {Seager}, Sara and {Sozzetti}, Alessandro and {Stassun}, Keivan G. and {Sullivan}, Peter and {Szentgyorgyi}, Andrew and {Torres}, Guillermo and {Udry}, Stephane and {Villasenor}, Joel},
        title = "{Transiting Exoplanet Survey Satellite (TESS)}",
      journal = {Journal of Astronomical Telescopes, Instruments, and Systems},
         year = 2015,
        month = jan,
       volume = {1},
          eid = {014003},
        pages = {014003},
          doi = {10.1117/1.JATIS.1.1.014003},
       adsurl = {https://ui.adsabs.harvard.edu/abs/2015JATIS...1a4003R}
}

@ARTICLE{2021AJ....162..162B,
       author = {{Barnett}, Joseph W. and {Williams}, Kurtis A. and {B{\'e}dard}, A. and {Bolte}, Michael},
        title = "{The Initial-Final Mass Relation for Hydrogen-deficient White Dwarfs}",
      journal = {\aj},
         year = 2021,
        month = oct,
       volume = {162},
       number = {4},
          eid = {162},
        pages = {162},
          doi = {10.3847/1538-3881/ac1423},
archivePrefix = {arXiv},
       eprint = {2107.06373},
 primaryClass = {astro-ph.SR},
       adsurl = {https://ui.adsabs.harvard.edu/abs/2021AJ....162..162B}
}

@ARTICLE{2024MNRAS.527.3602C,
       author = {{Cunningham}, Tim and {Tremblay}, Pier-Emmanuel and {W. O'Brien}, Mairi},
        title = "{Initial-final mass relation from white dwarfs within 40 pc}",
      journal = {\mnras},
         year = 2024,
        month = jan,
       volume = {527},
       number = {2},
        pages = {3602-3611},
          doi = {10.1093/mnras/stad3275},
archivePrefix = {arXiv},
       eprint = {2310.15410},
 primaryClass = {astro-ph.SR},
       adsurl = {https://ui.adsabs.harvard.edu/abs/2024MNRAS.527.3602C}
}

@ARTICLE{2019ApJ...871..169G,
       author = {{Genest-Beaulieu}, C. and {Bergeron}, P.},
        title = "{A Comprehensive Spectroscopic and Photometric Analysis of DA and DB White Dwarfs from SDSS and Gaia}",
      journal = {\apj},
         year = 2019,
        month = feb,
       volume = {871},
       number = {2},
          eid = {169},
        pages = {169},
          doi = {10.3847/1538-4357/aafac6},
       adsurl = {https://ui.adsabs.harvard.edu/abs/2019ApJ...871..169G}
}

@ARTICLE{2023A&A...674A...1G,
       author = {{Gaia Collaboration} and {Vallenari}, A. and {Brown}, A.~G.~A. and {Prusti}, T. and {de Bruijne}, J.~H.~J. and {Arenou}, F. and {Babusiaux}, C. and {Biermann}, M. and {Creevey}, O.~L. and {Ducourant}, C. and {Evans}, D.~W. and {Eyer}, L. and {Guerra}, R. and {Hutton}, A. and {Jordi}, C. and {Klioner}, S.~A. and {Lammers}, U.~L. and {Lindegren}, L. and {Luri}, X. and {Mignard}, F. and {Panem}, C. and {Pourbaix}, D. and {Randich}, S. and {Sartoretti}, P. and {Soubiran}, C. and {Tanga}, P. and {Walton}, N.~A. and {Bailer-Jones}, C.~A.~L. and {Bastian}, U. and {Drimmel}, R. and {Jansen}, F. and {Katz}, D. and {Lattanzi}, M.~G. and {van Leeuwen}, F. and {Bakker}, J. and {Cacciari}, C. and {Casta{\~n}eda}, J. and {De Angeli}, F. and {Fabricius}, C. and {Fouesneau}, M. and {Fr{\'e}mat}, Y. and {Galluccio}, L. and {Guerrier}, A. and {Heiter}, U. and {Masana}, E. and {Messineo}, R. and {Mowlavi}, N. and {Nicolas}, C. and {Nienartowicz}, K. and {Pailler}, F. and {Panuzzo}, P. and {Riclet}, F. and {Roux}, W. and {Seabroke}, G.~M. and {Sordo}, R. and {Th{\'e}venin}, F. and {Gracia-Abril}, G. and {Portell}, J. and {Teyssier}, D. and {Altmann}, M. and {Andrae}, R. and {Audard}, M. and {Bellas-Velidis}, I. and {Benson}, K. and {Berthier}, J. and {Blomme}, R. and {Burgess}, P.~W. and {Busonero}, D. and {Busso}, G. and {C{\'a}novas}, H. and {Carry}, B. and {Cellino}, A. and {Cheek}, N. and {Clementini}, G. and {Damerdji}, Y. and {Davidson}, M. and {de Teodoro}, P. and {Nu{\~n}ez Campos}, M. and {Delchambre}, L. and {Dell'Oro}, A. and {Esquej}, P. and {Fern{\'a}ndez-Hern{\'a}ndez}, J. and {Fraile}, E. and {Garabato}, D. and {Garc{\'\i}a-Lario}, P. and {Gosset}, E. and {Haigron}, R. and {Halbwachs}, J. -L. and {Hambly}, N.~C. and {Harrison}, D.~L. and {Hern{\'a}ndez}, J. and {Hestroffer}, D. and {Hodgkin}, S.~T. and {Holl}, B. and {Jan{\ss}en}, K. and {Jevardat de Fombelle}, G. and {Jordan}, S. and {Krone-Martins}, A. and {Lanzafame}, A.~C. and {L{\"o}ffler}, W. and {Marchal}, O. and {Marrese}, P.~M. and {Moitinho}, A. and {Muinonen}, K. and {Osborne}, P. and {Pancino}, E. and {Pauwels}, T. and {Recio-Blanco}, A. and {Reyl{\'e}}, C. and {Riello}, M. and {Rimoldini}, L. and {Roegiers}, T. and {Rybizki}, J. and {Sarro}, L.~M. and {Siopis}, C. and {Smith}, M. and {Sozzetti}, A. and {Utrilla}, E. and {van Leeuwen}, M. and {Abbas}, U. and {{\'A}brah{\'a}m}, P. and {Abreu Aramburu}, A. and {Aerts}, C. and {Aguado}, J.~J. and {Ajaj}, M. and {Aldea-Montero}, F. and {Altavilla}, G. and {{\'A}lvarez}, M.~A. and {Alves}, J. and {Anders}, F. and {Anderson}, R.~I. and {Anglada Varela}, E. and {Antoja}, T. and {Baines}, D. and {Baker}, S.~G. and {Balaguer-N{\'u}{\~n}ez}, L. and {Balbinot}, E. and {Balog}, Z. and {Barache}, C. and {Barbato}, D. and {Barros}, M. and {Barstow}, M.~A. and {Bartolom{\'e}}, S. and {Bassilana}, J. -L. and {Bauchet}, N. and {Becciani}, U. and {Bellazzini}, M. and {Berihuete}, A. and {Bernet}, M. and {Bertone}, S. and {Bianchi}, L. and {Binnenfeld}, A. and {Blanco-Cuaresma}, S. and {Blazere}, A. and {Boch}, T. and {Bombrun}, A. and {Bossini}, D. and {Bouquillon}, S. and {Bragaglia}, A. and {Bramante}, L. and {Breedt}, E. and {Bressan}, A. and {Brouillet}, N. and {Brugaletta}, E. and {Bucciarelli}, B. and {Burlacu}, A. and {Butkevich}, A.~G. and {Buzzi}, R. and {Caffau}, E. and {Cancelliere}, R. and {Cantat-Gaudin}, T. and {Carballo}, R. and {Carlucci}, T. and {Carnerero}, M.~I. and {Carrasco}, J.~M. and {Casamiquela}, L. and {Castellani}, M. and {Castro-Ginard}, A. and {Chaoul}, L. and {Charlot}, P. and {Chemin}, L. and {Chiaramida}, V. and {Chiavassa}, A. and {Chornay}, N. and {Comoretto}, G. and {Contursi}, G. and {Cooper}, W.~J. and {Cornez}, T. and {Cowell}, S. and {Crifo}, F. and {Cropper}, M. and {Crosta}, M. and {Crowley}, C. and {Dafonte}, C. and {Dapergolas}, A. and {David}, M. and {David}, P. and {de Laverny}, P. and {De Luise}, F. and {De March}, R. and {De Ridder}, J. and {de Souza}, R. and {de Torres}, A. and {del Peloso}, E.~F. and {del Pozo}, E. and {Delbo}, M. and {Delgado}, A. and {Delisle}, J. -B. and {Demouchy}, C. and {Dharmawardena}, T.~E. and {Di Matteo}, P. and {Diakite}, S. and {Diener}, C. and {Distefano}, E. and {Dolding}, C. and {Edvardsson}, B. and {Enke}, H. and {Fabre}, C. and {Fabrizio}, M. and {Faigler}, S. and {Fedorets}, G. and {Fernique}, P. and {Fienga}, A. and {Figueras}, F. and {Fournier}, Y. and {Fouron}, C. and {Fragkoudi}, F. and {Gai}, M. and {Garcia-Gutierrez}, A. and {Garcia-Reinaldos}, M. and {Garc{\'\i}a-Torres}, M. and {Garofalo}, A. and {Gavel}, A. and {Gavras}, P. and {Gerlach}, E. and {Geyer}, R. and {Giacobbe}, P. and {Gilmore}, G. and {Girona}, S. and {Giuffrida}, G. and {Gomel}, R. and {Gomez}, A. and {Gonz{\'a}lez-N{\'u}{\~n}ez}, J. and {Gonz{\'a}lez-Santamar{\'\i}a}, I. and {Gonz{\'a}lez-Vidal}, J.~J. and {Granvik}, M. and {Guillout}, P. and {Guiraud}, J. and {Guti{\'e}rrez-S{\'a}nchez}, R. and {Guy}, L.~P. and {Hatzidimitriou}, D. and {Hauser}, M. and {Haywood}, M. and {Helmer}, A. and {Helmi}, A. and {Sarmiento}, M.~H. and {Hidalgo}, S.~L. and {Hilger}, T. and {H{\l}adczuk}, N. and {Hobbs}, D. and {Holland}, G. and {Huckle}, H.~E. and {Jardine}, K. and {Jasniewicz}, G. and {Jean-Antoine Piccolo}, A. and {Jim{\'e}nez-Arranz}, {\'O}. and {Jorissen}, A. and {Juaristi Campillo}, J. and {Julbe}, F. and {Karbevska}, L. and {Kervella}, P. and {Khanna}, S. and {Kontizas}, M. and {Kordopatis}, G. and {Korn}, A.~J. and {K{\'o}sp{\'a}l}, {\'A}. and {Kostrzewa-Rutkowska}, Z. and {Kruszy{\'n}ska}, K. and {Kun}, M. and {Laizeau}, P. and {Lambert}, S. and {Lanza}, A.~F. and {Lasne}, Y. and {Le Campion}, J. -F. and {Lebreton}, Y. and {Lebzelter}, T. and {Leccia}, S. and {Leclerc}, N. and {Lecoeur-Taibi}, I. and {Liao}, S. and {Licata}, E.~L. and {Lindstr{\o}m}, H.~E.~P. and {Lister}, T.~A. and {Livanou}, E. and {Lobel}, A. and {Lorca}, A. and {Loup}, C. and {Madrero Pardo}, P. and {Magdaleno Romeo}, A. and {Managau}, S. and {Mann}, R.~G. and {Manteiga}, M. and {Marchant}, J.~M. and {Marconi}, M. and {Marcos}, J. and {Marcos Santos}, M.~M.~S. and {Mar{\'\i}n Pina}, D. and {Marinoni}, S. and {Marocco}, F. and {Marshall}, D.~J. and {Martin Polo}, L. and {Mart{\'\i}n-Fleitas}, J.~M. and {Marton}, G. and {Mary}, N. and {Masip}, A. and {Massari}, D. and {Mastrobuono-Battisti}, A. and {Mazeh}, T. and {McMillan}, P.~J. and {Messina}, S. and {Michalik}, D. and {Millar}, N.~R. and {Mints}, A. and {Molina}, D. and {Molinaro}, R. and {Moln{\'a}r}, L. and {Monari}, G. and {Mongui{\'o}}, M. and {Montegriffo}, P. and {Montero}, A. and {Mor}, R. and {Mora}, A. and {Morbidelli}, R. and {Morel}, T. and {Morris}, D. and {Muraveva}, T. and {Murphy}, C.~P. and {Musella}, I. and {Nagy}, Z. and {Noval}, L. and {Oca{\~n}a}, F. and {Ogden}, A. and {Ordenovic}, C. and {Osinde}, J.~O. and {Pagani}, C. and {Pagano}, I. and {Palaversa}, L. and {Palicio}, P.~A. and {Pallas-Quintela}, L. and {Panahi}, A. and {Payne-Wardenaar}, S. and {Pe{\~n}alosa Esteller}, X. and {Penttil{\"a}}, A. and {Pichon}, B. and {Piersimoni}, A.~M. and {Pineau}, F. -X. and {Plachy}, E. and {Plum}, G. and {Poggio}, E. and {Pr{\v{s}}a}, A. and {Pulone}, L. and {Racero}, E. and {Ragaini}, S. and {Rainer}, M. and {Raiteri}, C.~M. and {Rambaux}, N. and {Ramos}, P. and {Ramos-Lerate}, M. and {Re Fiorentin}, P. and {Regibo}, S. and {Richards}, P.~J. and {Rios Diaz}, C. and {Ripepi}, V. and {Riva}, A. and {Rix}, H. -W. and {Rixon}, G. and {Robichon}, N. and {Robin}, A.~C. and {Robin}, C. and {Roelens}, M. and {Rogues}, H.~R.~O. and {Rohrbasser}, L. and {Romero-G{\'o}mez}, M. and {Rowell}, N. and {Royer}, F. and {Ruz Mieres}, D. and {Rybicki}, K.~A. and {Sadowski}, G. and {S{\'a}ez N{\'u}{\~n}ez}, A. and {Sagrist{\`a} Sell{\'e}s}, A. and {Sahlmann}, J. and {Salguero}, E. and {Samaras}, N. and {Sanchez Gimenez}, V. and {Sanna}, N. and {Santove{\~n}a}, R. and {Sarasso}, M. and {Schultheis}, M. and {Sciacca}, E. and {Segol}, M. and {Segovia}, J.~C. and {S{\'e}gransan}, D. and {Semeux}, D. and {Shahaf}, S. and {Siddiqui}, H.~I. and {Siebert}, A. and {Siltala}, L. and {Silvelo}, A. and {Slezak}, E. and {Slezak}, I. and {Smart}, R.~L. and {Snaith}, O.~N. and {Solano}, E. and {Solitro}, F. and {Souami}, D. and {Souchay}, J. and {Spagna}, A. and {Spina}, L. and {Spoto}, F. and {Steele}, I.~A. and {Steidelm{\"u}ller}, H. and {Stephenson}, C.~A. and {S{\"u}veges}, M. and {Surdej}, J. and {Szabados}, L. and {Szegedi-Elek}, E. and {Taris}, F. and {Taylor}, M.~B. and {Teixeira}, R. and {Tolomei}, L. and {Tonello}, N. and {Torra}, F. and {Torra}, J. and {Torralba Elipe}, G. and {Trabucchi}, M. and {Tsounis}, A.~T. and {Turon}, C. and {Ulla}, A. and {Unger}, N. and {Vaillant}, M.~V. and {van Dillen}, E. and {van Reeven}, W. and {Vanel}, O. and {Vecchiato}, A. and {Viala}, Y. and {Vicente}, D. and {Voutsinas}, S. and {Weiler}, M. and {Wevers}, T. and {Wyrzykowski}, {\L}. and {Yoldas}, A. and {Yvard}, P. and {Zhao}, H. and {Zorec}, J. and {Zucker}, S. and {Zwitter}, T.},
        title = "{Gaia Data Release 3. Summary of the content and survey properties}",
      journal = {\aap},
         year = 2023,
        month = jun,
       volume = {674},
          eid = {A1},
        pages = {A1},
          doi = {10.1051/0004-6361/202243940},
archivePrefix = {arXiv},
       eprint = {2208.00211},
 primaryClass = {astro-ph.GA},
       adsurl = {https://ui.adsabs.harvard.edu/abs/2023A&A...674A...1G}
}

@ARTICLE{2018A&A...616A...1G,
       author = {{Gaia Collaboration} and {Brown}, A.~G.~A. and {Vallenari}, A. and {Prusti}, T. and {de Bruijne}, J.~H.~J. and {Babusiaux}, C. and {Bailer-Jones}, C.~A.~L. and {Biermann}, M. and {Evans}, D.~W. and {Eyer}, L. and {Jansen}, F. and {Jordi}, C. and {Klioner}, S.~A. and {Lammers}, U. and {Lindegren}, L. and {Luri}, X. and {Mignard}, F. and {Panem}, C. and {Pourbaix}, D. and {Randich}, S. and {Sartoretti}, P. and {Siddiqui}, H.~I. and {Soubiran}, C. and {van Leeuwen}, F. and {Walton}, N.~A. and {Arenou}, F. and {Bastian}, U. and {Cropper}, M. and {Drimmel}, R. and {Katz}, D. and {Lattanzi}, M.~G. and {Bakker}, J. and {Cacciari}, C. and {Casta{\~n}eda}, J. and {Chaoul}, L. and {Cheek}, N. and {De Angeli}, F. and {Fabricius}, C. and {Guerra}, R. and {Holl}, B. and {Masana}, E. and {Messineo}, R. and {Mowlavi}, N. and {Nienartowicz}, K. and {Panuzzo}, P. and {Portell}, J. and {Riello}, M. and {Seabroke}, G.~M. and {Tanga}, P. and {Th{\'e}venin}, F. and {Gracia-Abril}, G. and {Comoretto}, G. and {Garcia-Reinaldos}, M. and {Teyssier}, D. and {Altmann}, M. and {Andrae}, R. and {Audard}, M. and {Bellas-Velidis}, I. and {Benson}, K. and {Berthier}, J. and {Blomme}, R. and {Burgess}, P. and {Busso}, G. and {Carry}, B. and {Cellino}, A. and {Clementini}, G. and {Clotet}, M. and {Creevey}, O. and {Davidson}, M. and {De Ridder}, J. and {Delchambre}, L. and {Dell'Oro}, A. and {Ducourant}, C. and {Fern{\'a}ndez-Hern{\'a}ndez}, J. and {Fouesneau}, M. and {Fr{\'e}mat}, Y. and {Galluccio}, L. and {Garc{\'\i}a-Torres}, M. and {Gonz{\'a}lez-N{\'u}{\~n}ez}, J. and {Gonz{\'a}lez-Vidal}, J.~J. and {Gosset}, E. and {Guy}, L.~P. and {Halbwachs}, J. -L. and {Hambly}, N.~C. and {Harrison}, D.~L. and {Hern{\'a}ndez}, J. and {Hestroffer}, D. and {Hodgkin}, S.~T. and {Hutton}, A. and {Jasniewicz}, G. and {Jean-Antoine-Piccolo}, A. and {Jordan}, S. and {Korn}, A.~J. and {Krone-Martins}, A. and {Lanzafame}, A.~C. and {Lebzelter}, T. and {L{\"o}ffler}, W. and {Manteiga}, M. and {Marrese}, P.~M. and {Mart{\'\i}n-Fleitas}, J.~M. and {Moitinho}, A. and {Mora}, A. and {Muinonen}, K. and {Osinde}, J. and {Pancino}, E. and {Pauwels}, T. and {Petit}, J. -M. and {Recio-Blanco}, A. and {Richards}, P.~J. and {Rimoldini}, L. and {Robin}, A.~C. and {Sarro}, L.~M. and {Siopis}, C. and {Smith}, M. and {Sozzetti}, A. and {S{\"u}veges}, M. and {Torra}, J. and {van Reeven}, W. and {Abbas}, U. and {Abreu Aramburu}, A. and {Accart}, S. and {Aerts}, C. and {Altavilla}, G. and {{\'A}lvarez}, M.~A. and {Alvarez}, R. and {Alves}, J. and {Anderson}, R.~I. and {Andrei}, A.~H. and {Anglada Varela}, E. and {Antiche}, E. and {Antoja}, T. and {Arcay}, B. and {Astraatmadja}, T.~L. and {Bach}, N. and {Baker}, S.~G. and {Balaguer-N{\'u}{\~n}ez}, L. and {Balm}, P. and {Barache}, C. and {Barata}, C. and {Barbato}, D. and {Barblan}, F. and {Barklem}, P.~S. and {Barrado}, D. and {Barros}, M. and {Barstow}, M.~A. and {Bartholom{\'e} Mu{\~n}oz}, S. and {Bassilana}, J. -L. and {Becciani}, U. and {Bellazzini}, M. and {Berihuete}, A. and {Bertone}, S. and {Bianchi}, L. and {Bienaym{\'e}}, O. and {Blanco-Cuaresma}, S. and {Boch}, T. and {Boeche}, C. and {Bombrun}, A. and {Borrachero}, R. and {Bossini}, D. and {Bouquillon}, S. and {Bourda}, G. and {Bragaglia}, A. and {Bramante}, L. and {Breddels}, M.~A. and {Bressan}, A. and {Brouillet}, N. and {Br{\"u}semeister}, T. and {Brugaletta}, E. and {Bucciarelli}, B. and {Burlacu}, A. and {Busonero}, D. and {Butkevich}, A.~G. and {Buzzi}, R. and {Caffau}, E. and {Cancelliere}, R. and {Cannizzaro}, G. and {Cantat-Gaudin}, T. and {Carballo}, R. and {Carlucci}, T. and {Carrasco}, J.~M. and {Casamiquela}, L. and {Castellani}, M. and {Castro-Ginard}, A. and {Charlot}, P. and {Chemin}, L. and {Chiavassa}, A. and {Cocozza}, G. and {Costigan}, G. and {Cowell}, S. and {Crifo}, F. and {Crosta}, M. and {Crowley}, C. and {Cuypers}, J. and {Dafonte}, C. and {Damerdji}, Y. and {Dapergolas}, A. and {David}, P. and {David}, M. and {de Laverny}, P. and {De Luise}, F. and {De March}, R. and {de Martino}, D. and {de Souza}, R. and {de Torres}, A. and {Debosscher}, J. and {del Pozo}, E. and {Delbo}, M. and {Delgado}, A. and {Delgado}, H.~E. and {Di Matteo}, P. and {Diakite}, S. and {Diener}, C. and {Distefano}, E. and {Dolding}, C. and {Drazinos}, P. and {Dur{\'a}n}, J. and {Edvardsson}, B. and {Enke}, H. and {Eriksson}, K. and {Esquej}, P. and {Eynard Bontemps}, G. and {Fabre}, C. and {Fabrizio}, M. and {Faigler}, S. and {Falc{\~a}o}, A.~J. and {Farr{\`a}s Casas}, M. and {Federici}, L. and {Fedorets}, G. and {Fernique}, P. and {Figueras}, F. and {Filippi}, F. and {Findeisen}, K. and {Fonti}, A. and {Fraile}, E. and {Fraser}, M. and {Fr{\'e}zouls}, B. and {Gai}, M. and {Galleti}, S. and {Garabato}, D. and {Garc{\'\i}a-Sedano}, F. and {Garofalo}, A. and {Garralda}, N. and {Gavel}, A. and {Gavras}, P. and {Gerssen}, J. and {Geyer}, R. and {Giacobbe}, P. and {Gilmore}, G. and {Girona}, S. and {Giuffrida}, G. and {Glass}, F. and {Gomes}, M. and {Granvik}, M. and {Gueguen}, A. and {Guerrier}, A. and {Guiraud}, J. and {Guti{\'e}rrez-S{\'a}nchez}, R. and {Haigron}, R. and {Hatzidimitriou}, D. and {Hauser}, M. and {Haywood}, M. and {Heiter}, U. and {Helmi}, A. and {Heu}, J. and {Hilger}, T. and {Hobbs}, D. and {Hofmann}, W. and {Holland}, G. and {Huckle}, H.~E. and {Hypki}, A. and {Icardi}, V. and {Jan{\ss}en}, K. and {Jevardat de Fombelle}, G. and {Jonker}, P.~G. and {Juh{\'a}sz}, {\'A}. L. and {Julbe}, F. and {Karampelas}, A. and {Kewley}, A. and {Klar}, J. and {Kochoska}, A. and {Kohley}, R. and {Kolenberg}, K. and {Kontizas}, M. and {Kontizas}, E. and {Koposov}, S.~E. and {Kordopatis}, G. and {Kostrzewa-Rutkowska}, Z. and {Koubsky}, P. and {Lambert}, S. and {Lanza}, A.~F. and {Lasne}, Y. and {Lavigne}, J. -B. and {Le Fustec}, Y. and {Le Poncin-Lafitte}, C. and {Lebreton}, Y. and {Leccia}, S. and {Leclerc}, N. and {Lecoeur-Taibi}, I. and {Lenhardt}, H. and {Leroux}, F. and {Liao}, S. and {Licata}, E. and {Lindstr{\o}m}, H.~E.~P. and {Lister}, T.~A. and {Livanou}, E. and {Lobel}, A. and {L{\'o}pez}, M. and {Managau}, S. and {Mann}, R.~G. and {Mantelet}, G. and {Marchal}, O. and {Marchant}, J.~M. and {Marconi}, M. and {Marinoni}, S. and {Marschalk{\'o}}, G. and {Marshall}, D.~J. and {Martino}, M. and {Marton}, G. and {Mary}, N. and {Massari}, D. and {Matijevi{\v{c}}}, G. and {Mazeh}, T. and {McMillan}, P.~J. and {Messina}, S. and {Michalik}, D. and {Millar}, N.~R. and {Molina}, D. and {Molinaro}, R. and {Moln{\'a}r}, L. and {Montegriffo}, P. and {Mor}, R. and {Morbidelli}, R. and {Morel}, T. and {Morris}, D. and {Mulone}, A.~F. and {Muraveva}, T. and {Musella}, I. and {Nelemans}, G. and {Nicastro}, L. and {Noval}, L. and {O'Mullane}, W. and {Ord{\'e}novic}, C. and {Ord{\'o}{\~n}ez-Blanco}, D. and {Osborne}, P. and {Pagani}, C. and {Pagano}, I. and {Pailler}, F. and {Palacin}, H. and {Palaversa}, L. and {Panahi}, A. and {Pawlak}, M. and {Piersimoni}, A.~M. and {Pineau}, F. -X. and {Plachy}, E. and {Plum}, G. and {Poggio}, E. and {Poujoulet}, E. and {Pr{\v{s}}a}, A. and {Pulone}, L. and {Racero}, E. and {Ragaini}, S. and {Rambaux}, N. and {Ramos-Lerate}, M. and {Regibo}, S. and {Reyl{\'e}}, C. and {Riclet}, F. and {Ripepi}, V. and {Riva}, A. and {Rivard}, A. and {Rixon}, G. and {Roegiers}, T. and {Roelens}, M. and {Romero-G{\'o}mez}, M. and {Rowell}, N. and {Royer}, F. and {Ruiz-Dern}, L. and {Sadowski}, G. and {Sagrist{\`a} Sell{\'e}s}, T. and {Sahlmann}, J. and {Salgado}, J. and {Salguero}, E. and {Sanna}, N. and {Santana-Ros}, T. and {Sarasso}, M. and {Savietto}, H. and {Schultheis}, M. and {Sciacca}, E. and {Segol}, M. and {Segovia}, J.~C. and {S{\'e}gransan}, D. and {Shih}, I. -C. and {Siltala}, L. and {Silva}, A.~F. and {Smart}, R.~L. and {Smith}, K.~W. and {Solano}, E. and {Solitro}, F. and {Sordo}, R. and {Soria Nieto}, S. and {Souchay}, J. and {Spagna}, A. and {Spoto}, F. and {Stampa}, U. and {Steele}, I.~A. and {Steidelm{\"u}ller}, H. and {Stephenson}, C.~A. and {Stoev}, H. and {Suess}, F.~F. and {Surdej}, J. and {Szabados}, L. and {Szegedi-Elek}, E. and {Tapiador}, D. and {Taris}, F. and {Tauran}, G. and {Taylor}, M.~B. and {Teixeira}, R. and {Terrett}, D. and {Teyssandier}, P. and {Thuillot}, W. and {Titarenko}, A. and {Torra Clotet}, F. and {Turon}, C. and {Ulla}, A. and {Utrilla}, E. and {Uzzi}, S. and {Vaillant}, M. and {Valentini}, G. and {Valette}, V. and {van Elteren}, A. and {Van Hemelryck}, E. and {van Leeuwen}, M. and {Vaschetto}, M. and {Vecchiato}, A. and {Veljanoski}, J. and {Viala}, Y. and {Vicente}, D. and {Vogt}, S. and {von Essen}, C. and {Voss}, H. and {Votruba}, V. and {Voutsinas}, S. and {Walmsley}, G. and {Weiler}, M. and {Wertz}, O. and {Wevers}, T. and {Wyrzykowski}, {\L}. and {Yoldas}, A. and {{\v{Z}}erjal}, M. and {Ziaeepour}, H. and {Zorec}, J. and {Zschocke}, S. and {Zucker}, S. and {Zurbach}, C. and {Zwitter}, T.},
        title = "{Gaia Data Release 2. Summary of the contents and survey properties}",
      journal = {\aap},
         year = 2018,
        month = aug,
       volume = {616},
          eid = {A1},
        pages = {A1},
          doi = {10.1051/0004-6361/201833051},
archivePrefix = {arXiv},
       eprint = {1804.09365},
 primaryClass = {astro-ph.GA},
       adsurl = {https://ui.adsabs.harvard.edu/abs/2018A&A...616A...1G}
}

@ARTICLE{2016A&A...595A...2G,
       author = {{Gaia Collaboration} and {Brown}, A.~G.~A. and {Vallenari}, A. and {Prusti}, T. and {de Bruijne}, J.~H.~J. and {Mignard}, F. and {Drimmel}, R. and {Babusiaux}, C. and {Bailer-Jones}, C.~A.~L. and {Bastian}, U. and {Biermann}, M. and {Evans}, D.~W. and {Eyer}, L. and {Jansen}, F. and {Jordi}, C. and {Katz}, D. and {Klioner}, S.~A. and {Lammers}, U. and {Lindegren}, L. and {Luri}, X. and {O'Mullane}, W. and {Panem}, C. and {Pourbaix}, D. and {Randich}, S. and {Sartoretti}, P. and {Siddiqui}, H.~I. and {Soubiran}, C. and {Valette}, V. and {van Leeuwen}, F. and {Walton}, N.~A. and {Aerts}, C. and {Arenou}, F. and {Cropper}, M. and {H{\o}g}, E. and {Lattanzi}, M.~G. and {Grebel}, E.~K. and {Holland}, A.~D. and {Huc}, C. and {Passot}, X. and {Perryman}, M. and {Bramante}, L. and {Cacciari}, C. and {Casta{\~n}eda}, J. and {Chaoul}, L. and {Cheek}, N. and {De Angeli}, F. and {Fabricius}, C. and {Guerra}, R. and {Hern{\'a}ndez}, J. and {Jean-Antoine-Piccolo}, A. and {Masana}, E. and {Messineo}, R. and {Mowlavi}, N. and {Nienartowicz}, K. and {Ord{\'o}{\~n}ez-Blanco}, D. and {Panuzzo}, P. and {Portell}, J. and {Richards}, P.~J. and {Riello}, M. and {Seabroke}, G.~M. and {Tanga}, P. and {Th{\'e}venin}, F. and {Torra}, J. and {Els}, S.~G. and {Gracia-Abril}, G. and {Comoretto}, G. and {Garcia-Reinaldos}, M. and {Lock}, T. and {Mercier}, E. and {Altmann}, M. and {Andrae}, R. and {Astraatmadja}, T.~L. and {Bellas-Velidis}, I. and {Benson}, K. and {Berthier}, J. and {Blomme}, R. and {Busso}, G. and {Carry}, B. and {Cellino}, A. and {Clementini}, G. and {Cowell}, S. and {Creevey}, O. and {Cuypers}, J. and {Davidson}, M. and {De Ridder}, J. and {de Torres}, A. and {Delchambre}, L. and {Dell'Oro}, A. and {Ducourant}, C. and {Fr{\'e}mat}, Y. and {Garc{\'\i}a-Torres}, M. and {Gosset}, E. and {Halbwachs}, J. -L. and {Hambly}, N.~C. and {Harrison}, D.~L. and {Hauser}, M. and {Hestroffer}, D. and {Hodgkin}, S.~T. and {Huckle}, H.~E. and {Hutton}, A. and {Jasniewicz}, G. and {Jordan}, S. and {Kontizas}, M. and {Korn}, A.~J. and {Lanzafame}, A.~C. and {Manteiga}, M. and {Moitinho}, A. and {Muinonen}, K. and {Osinde}, J. and {Pancino}, E. and {Pauwels}, T. and {Petit}, J. -M. and {Recio-Blanco}, A. and {Robin}, A.~C. and {Sarro}, L.~M. and {Siopis}, C. and {Smith}, M. and {Smith}, K.~W. and {Sozzetti}, A. and {Thuillot}, W. and {van Reeven}, W. and {Viala}, Y. and {Abbas}, U. and {Abreu Aramburu}, A. and {Accart}, S. and {Aguado}, J.~J. and {Allan}, P.~M. and {Allasia}, W. and {Altavilla}, G. and {{\'A}lvarez}, M.~A. and {Alves}, J. and {Anderson}, R.~I. and {Andrei}, A.~H. and {Anglada Varela}, E. and {Antiche}, E. and {Antoja}, T. and {Ant{\'o}n}, S. and {Arcay}, B. and {Bach}, N. and {Baker}, S.~G. and {Balaguer-N{\'u}{\~n}ez}, L. and {Barache}, C. and {Barata}, C. and {Barbier}, A. and {Barblan}, F. and {Barrado y Navascu{\'e}s}, D. and {Barros}, M. and {Barstow}, M.~A. and {Becciani}, U. and {Bellazzini}, M. and {Bello Garc{\'\i}a}, A. and {Belokurov}, V. and {Bendjoya}, P. and {Berihuete}, A. and {Bianchi}, L. and {Bienaym{\'e}}, O. and {Billebaud}, F. and {Blagorodnova}, N. and {Blanco-Cuaresma}, S. and {Boch}, T. and {Bombrun}, A. and {Borrachero}, R. and {Bouquillon}, S. and {Bourda}, G. and {Bouy}, H. and {Bragaglia}, A. and {Breddels}, M.~A. and {Brouillet}, N. and {Br{\"u}semeister}, T. and {Bucciarelli}, B. and {Burgess}, P. and {Burgon}, R. and {Burlacu}, A. and {Busonero}, D. and {Buzzi}, R. and {Caffau}, E. and {Cambras}, J. and {Campbell}, H. and {Cancelliere}, R. and {Cantat-Gaudin}, T. and {Carlucci}, T. and {Carrasco}, J.~M. and {Castellani}, M. and {Charlot}, P. and {Charnas}, J. and {Chiavassa}, A. and {Clotet}, M. and {Cocozza}, G. and {Collins}, R.~S. and {Costigan}, G. and {Crifo}, F. and {Cross}, N.~J.~G. and {Crosta}, M. and {Crowley}, C. and {Dafonte}, C. and {Damerdji}, Y. and {Dapergolas}, A. and {David}, P. and {David}, M. and {De Cat}, P. and {de Felice}, F. and {de Laverny}, P. and {De Luise}, F. and {De March}, R. and {de Martino}, D. and {de Souza}, R. and {Debosscher}, J. and {del Pozo}, E. and {Delbo}, M. and {Delgado}, A. and {Delgado}, H.~E. and {Di Matteo}, P. and {Diakite}, S. and {Distefano}, E. and {Dolding}, C. and {Dos Anjos}, S. and {Drazinos}, P. and {Duran}, J. and {Dzigan}, Y. and {Edvardsson}, B. and {Enke}, H. and {Evans}, N.~W. and {Eynard Bontemps}, G. and {Fabre}, C. and {Fabrizio}, M. and {Faigler}, S. and {Falc{\~a}o}, A.~J. and {Farr{\`a}s Casas}, M. and {Federici}, L. and {Fedorets}, G. and {Fern{\'a}ndez-Hern{\'a}ndez}, J. and {Fernique}, P. and {Fienga}, A. and {Figueras}, F. and {Filippi}, F. and {Findeisen}, K. and {Fonti}, A. and {Fouesneau}, M. and {Fraile}, E. and {Fraser}, M. and {Fuchs}, J. and {Gai}, M. and {Galleti}, S. and {Galluccio}, L. and {Garabato}, D. and {Garc{\'\i}a-Sedano}, F. and {Garofalo}, A. and {Garralda}, N. and {Gavras}, P. and {Gerssen}, J. and {Geyer}, R. and {Gilmore}, G. and {Girona}, S. and {Giuffrida}, G. and {Gomes}, M. and {Gonz{\'a}lez-Marcos}, A. and {Gonz{\'a}lez-N{\'u}{\~n}ez}, J. and {Gonz{\'a}lez-Vidal}, J.~J. and {Granvik}, M. and {Guerrier}, A. and {Guillout}, P. and {Guiraud}, J. and {G{\'u}rpide}, A. and {Guti{\'e}rrez-S{\'a}nchez}, R. and {Guy}, L.~P. and {Haigron}, R. and {Hatzidimitriou}, D. and {Haywood}, M. and {Heiter}, U. and {Helmi}, A. and {Hobbs}, D. and {Hofmann}, W. and {Holl}, B. and {Holland}, G. and {Hunt}, J.~A.~S. and {Hypki}, A. and {Icardi}, V. and {Irwin}, M. and {Jevardat de Fombelle}, G. and {Jofr{\'e}}, P. and {Jonker}, P.~G. and {Jorissen}, A. and {Julbe}, F. and {Karampelas}, A. and {Kochoska}, A. and {Kohley}, R. and {Kolenberg}, K. and {Kontizas}, E. and {Koposov}, S.~E. and {Kordopatis}, G. and {Koubsky}, P. and {Krone-Martins}, A. and {Kudryashova}, M. and {Kull}, I. and {Bachchan}, R.~K. and {Lacoste-Seris}, F. and {Lanza}, A.~F. and {Lavigne}, J. -B. and {Le Poncin-Lafitte}, C. and {Lebreton}, Y. and {Lebzelter}, T. and {Leccia}, S. and {Leclerc}, N. and {Lecoeur-Taibi}, I. and {Lemaitre}, V. and {Lenhardt}, H. and {Leroux}, F. and {Liao}, S. and {Licata}, E. and {Lindstr{\o}m}, H.~E.~P. and {Lister}, T.~A. and {Livanou}, E. and {Lobel}, A. and {L{\"o}ffler}, W. and {L{\'o}pez}, M. and {Lorenz}, D. and {MacDonald}, I. and {Magalh{\~a}es Fernandes}, T. and {Managau}, S. and {Mann}, R.~G. and {Mantelet}, G. and {Marchal}, O. and {Marchant}, J.~M. and {Marconi}, M. and {Marinoni}, S. and {Marrese}, P.~M. and {Marschalk{\'o}}, G. and {Marshall}, D.~J. and {Mart{\'\i}n-Fleitas}, J.~M. and {Martino}, M. and {Mary}, N. and {Matijevi{\v{c}}}, G. and {Mazeh}, T. and {McMillan}, P.~J. and {Messina}, S. and {Michalik}, D. and {Millar}, N.~R. and {Miranda}, B.~M.~H. and {Molina}, D. and {Molinaro}, R. and {Molinaro}, M. and {Moln{\'a}r}, L. and {Moniez}, M. and {Montegriffo}, P. and {Mor}, R. and {Mora}, A. and {Morbidelli}, R. and {Morel}, T. and {Morgenthaler}, S. and {Morris}, D. and {Mulone}, A.~F. and {Muraveva}, T. and {Musella}, I. and {Narbonne}, J. and {Nelemans}, G. and {Nicastro}, L. and {Noval}, L. and {Ord{\'e}novic}, C. and {Ordieres-Mer{\'e}}, J. and {Osborne}, P. and {Pagani}, C. and {Pagano}, I. and {Pailler}, F. and {Palacin}, H. and {Palaversa}, L. and {Parsons}, P. and {Pecoraro}, M. and {Pedrosa}, R. and {Pentik{\"a}inen}, H. and {Pichon}, B. and {Piersimoni}, A.~M. and {Pineau}, F. -X. and {Plachy}, E. and {Plum}, G. and {Poujoulet}, E. and {Pr{\v{s}}a}, A. and {Pulone}, L. and {Ragaini}, S. and {Rago}, S. and {Rambaux}, N. and {Ramos-Lerate}, M. and {Ranalli}, P. and {Rauw}, G. and {Read}, A. and {Regibo}, S. and {Reyl{\'e}}, C. and {Ribeiro}, R.~A. and {Rimoldini}, L. and {Ripepi}, V. and {Riva}, A. and {Rixon}, G. and {Roelens}, M. and {Romero-G{\'o}mez}, M. and {Rowell}, N. and {Royer}, F. and {Ruiz-Dern}, L. and {Sadowski}, G. and {Sagrist{\`a} Sell{\'e}s}, T. and {Sahlmann}, J. and {Salgado}, J. and {Salguero}, E. and {Sarasso}, M. and {Savietto}, H. and {Schultheis}, M. and {Sciacca}, E. and {Segol}, M. and {Segovia}, J.~C. and {Segransan}, D. and {Shih}, I. -C. and {Smareglia}, R. and {Smart}, R.~L. and {Solano}, E. and {Solitro}, F. and {Sordo}, R. and {Soria Nieto}, S. and {Souchay}, J. and {Spagna}, A. and {Spoto}, F. and {Stampa}, U. and {Steele}, I.~A. and {Steidelm{\"u}ller}, H. and {Stephenson}, C.~A. and {Stoev}, H. and {Suess}, F.~F. and {S{\"u}veges}, M. and {Surdej}, J. and {Szabados}, L. and {Szegedi-Elek}, E. and {Tapiador}, D. and {Taris}, F. and {Tauran}, G. and {Taylor}, M.~B. and {Teixeira}, R. and {Terrett}, D. and {Tingley}, B. and {Trager}, S.~C. and {Turon}, C. and {Ulla}, A. and {Utrilla}, E. and {Valentini}, G. and {van Elteren}, A. and {Van Hemelryck}, E. and {van Leeuwen}, M. and {Varadi}, M. and {Vecchiato}, A. and {Veljanoski}, J. and {Via}, T. and {Vicente}, D. and {Vogt}, S. and {Voss}, H. and {Votruba}, V. and {Voutsinas}, S. and {Walmsley}, G. and {Weiler}, M. and {Weingrill}, K. and {Wevers}, T. and {Wyrzykowski}, {\L}. and {Yoldas}, A. and {{\v{Z}}erjal}, M. and {Zucker}, S. and {Zurbach}, C. and {Zwitter}, T. and {Alecu}, A. and {Allen}, M. and {Allende Prieto}, C. and {Amorim}, A. and {Anglada-Escud{\'e}}, G. and {Arsenijevic}, V. and {Azaz}, S. and {Balm}, P. and {Beck}, M. and {Bernstein}, H. -H. and {Bigot}, L. and {Bijaoui}, A. and {Blasco}, C. and {Bonfigli}, M. and {Bono}, G. and {Boudreault}, S. and {Bressan}, A. and {Brown}, S. and {Brunet}, P. -M. and {Bunclark}, P. and {Buonanno}, R. and {Butkevich}, A.~G. and {Carret}, C. and {Carrion}, C. and {Chemin}, L. and {Ch{\'e}reau}, F. and {Corcione}, L. and {Darmigny}, E. and {de Boer}, K.~S. and {de Teodoro}, P. and {de Zeeuw}, P.~T. and {Delle Luche}, C. and {Domingues}, C.~D. and {Dubath}, P. and {Fodor}, F. and {Fr{\'e}zouls}, B. and {Fries}, A. and {Fustes}, D. and {Fyfe}, D. and {Gallardo}, E. and {Gallegos}, J. and {Gardiol}, D. and {Gebran}, M. and {Gomboc}, A. and {G{\'o}mez}, A. and {Grux}, E. and {Gueguen}, A. and {Heyrovsky}, A. and {Hoar}, J. and {Iannicola}, G. and {Isasi Parache}, Y. and {Janotto}, A. -M. and {Joliet}, E. and {Jonckheere}, A. and {Keil}, R. and {Kim}, D. -W. and {Klagyivik}, P. and {Klar}, J. and {Knude}, J. and {Kochukhov}, O. and {Kolka}, I. and {Kos}, J. and {Kutka}, A. and {Lainey}, V. and {LeBouquin}, D. and {Liu}, C. and {Loreggia}, D. and {Makarov}, V.~V. and {Marseille}, M.~G. and {Martayan}, C. and {Martinez-Rubi}, O. and {Massart}, B. and {Meynadier}, F. and {Mignot}, S. and {Munari}, U. and {Nguyen}, A. -T. and {Nordlander}, T. and {Ocvirk}, P. and {O'Flaherty}, K.~S. and {Olias Sanz}, A. and {Ortiz}, P. and {Osorio}, J. and {Oszkiewicz}, D. and {Ouzounis}, A. and {Palmer}, M. and {Park}, P. and {Pasquato}, E. and {Peltzer}, C. and {Peralta}, J. and {P{\'e}turaud}, F. and {Pieniluoma}, T. and {Pigozzi}, E. and {Poels}, J. and {Prat}, G. and {Prod'homme}, T. and {Raison}, F. and {Rebordao}, J.~M. and {Risquez}, D. and {Rocca-Volmerange}, B. and {Rosen}, S. and {Ruiz-Fuertes}, M.~I. and {Russo}, F. and {Sembay}, S. and {Serraller Vizcaino}, I. and {Short}, A. and {Siebert}, A. and {Silva}, H. and {Sinachopoulos}, D. and {Slezak}, E. and {Soffel}, M. and {Sosnowska}, D. and {Strai{\v{z}}ys}, V. and {ter Linden}, M. and {Terrell}, D. and {Theil}, S. and {Tiede}, C. and {Troisi}, L. and {Tsalmantza}, P. and {Tur}, D. and {Vaccari}, M. and {Vachier}, F. and {Valles}, P. and {Van Hamme}, W. and {Veltz}, L. and {Virtanen}, J. and {Wallut}, J. -M. and {Wichmann}, R. and {Wilkinson}, M.~I. and {Ziaeepour}, H. and {Zschocke}, S.},
        title = "{Gaia Data Release 1. Summary of the astrometric, photometric, and survey properties}",
      journal = {\aap},
         year = 2016,
        month = nov,
       volume = {595},
          eid = {A2},
        pages = {A2},
          doi = {10.1051/0004-6361/201629512},
archivePrefix = {arXiv},
       eprint = {1609.04172},
 primaryClass = {astro-ph.IM},
       adsurl = {https://ui.adsabs.harvard.edu/abs/2016A&A...595A...2G}
}

@ARTICLE{2021A&A...649A...1G,
       author = {{Gaia Collaboration} and {Brown}, A.~G.~A. and {Vallenari}, A. and {Prusti}, T. and {de Bruijne}, J.~H.~J. and {Babusiaux}, C. and {Biermann}, M. and {Creevey}, O.~L. and {Evans}, D.~W. and {Eyer}, L. and {Hutton}, A. and {Jansen}, F. and {Jordi}, C. and {Klioner}, S.~A. and {Lammers}, U. and {Lindegren}, L. and {Luri}, X. and {Mignard}, F. and {Panem}, C. and {Pourbaix}, D. and {Randich}, S. and {Sartoretti}, P. and {Soubiran}, C. and {Walton}, N.~A. and {Arenou}, F. and {Bailer-Jones}, C.~A.~L. and {Bastian}, U. and {Cropper}, M. and {Drimmel}, R. and {Katz}, D. and {Lattanzi}, M.~G. and {van Leeuwen}, F. and {Bakker}, J. and {Cacciari}, C. and {Casta{\~n}eda}, J. and {De Angeli}, F. and {Ducourant}, C. and {Fabricius}, C. and {Fouesneau}, M. and {Fr{\'e}mat}, Y. and {Guerra}, R. and {Guerrier}, A. and {Guiraud}, J. and {Jean-Antoine Piccolo}, A. and {Masana}, E. and {Messineo}, R. and {Mowlavi}, N. and {Nicolas}, C. and {Nienartowicz}, K. and {Pailler}, F. and {Panuzzo}, P. and {Riclet}, F. and {Roux}, W. and {Seabroke}, G.~M. and {Sordo}, R. and {Tanga}, P. and {Th{\'e}venin}, F. and {Gracia-Abril}, G. and {Portell}, J. and {Teyssier}, D. and {Altmann}, M. and {Andrae}, R. and {Bellas-Velidis}, I. and {Benson}, K. and {Berthier}, J. and {Blomme}, R. and {Brugaletta}, E. and {Burgess}, P.~W. and {Busso}, G. and {Carry}, B. and {Cellino}, A. and {Cheek}, N. and {Clementini}, G. and {Damerdji}, Y. and {Davidson}, M. and {Delchambre}, L. and {Dell'Oro}, A. and {Fern{\'a}ndez-Hern{\'a}ndez}, J. and {Galluccio}, L. and {Garc{\'\i}a-Lario}, P. and {Garcia-Reinaldos}, M. and {Gonz{\'a}lez-N{\'u}{\~n}ez}, J. and {Gosset}, E. and {Haigron}, R. and {Halbwachs}, J. -L. and {Hambly}, N.~C. and {Harrison}, D.~L. and {Hatzidimitriou}, D. and {Heiter}, U. and {Hern{\'a}ndez}, J. and {Hestroffer}, D. and {Hodgkin}, S.~T. and {Holl}, B. and {Jan{\ss}en}, K. and {Jevardat de Fombelle}, G. and {Jordan}, S. and {Krone-Martins}, A. and {Lanzafame}, A.~C. and {L{\"o}ffler}, W. and {Lorca}, A. and {Manteiga}, M. and {Marchal}, O. and {Marrese}, P.~M. and {Moitinho}, A. and {Mora}, A. and {Muinonen}, K. and {Osborne}, P. and {Pancino}, E. and {Pauwels}, T. and {Petit}, J. -M. and {Recio-Blanco}, A. and {Richards}, P.~J. and {Riello}, M. and {Rimoldini}, L. and {Robin}, A.~C. and {Roegiers}, T. and {Rybizki}, J. and {Sarro}, L.~M. and {Siopis}, C. and {Smith}, M. and {Sozzetti}, A. and {Ulla}, A. and {Utrilla}, E. and {van Leeuwen}, M. and {van Reeven}, W. and {Abbas}, U. and {Abreu Aramburu}, A. and {Accart}, S. and {Aerts}, C. and {Aguado}, J.~J. and {Ajaj}, M. and {Altavilla}, G. and {{\'A}lvarez}, M.~A. and {{\'A}lvarez Cid-Fuentes}, J. and {Alves}, J. and {Anderson}, R.~I. and {Anglada Varela}, E. and {Antoja}, T. and {Audard}, M. and {Baines}, D. and {Baker}, S.~G. and {Balaguer-N{\'u}{\~n}ez}, L. and {Balbinot}, E. and {Balog}, Z. and {Barache}, C. and {Barbato}, D. and {Barros}, M. and {Barstow}, M.~A. and {Bartolom{\'e}}, S. and {Bassilana}, J. -L. and {Bauchet}, N. and {Baudesson-Stella}, A. and {Becciani}, U. and {Bellazzini}, M. and {Bernet}, M. and {Bertone}, S. and {Bianchi}, L. and {Blanco-Cuaresma}, S. and {Boch}, T. and {Bombrun}, A. and {Bossini}, D. and {Bouquillon}, S. and {Bragaglia}, A. and {Bramante}, L. and {Breedt}, E. and {Bressan}, A. and {Brouillet}, N. and {Bucciarelli}, B. and {Burlacu}, A. and {Busonero}, D. and {Butkevich}, A.~G. and {Buzzi}, R. and {Caffau}, E. and {Cancelliere}, R. and {C{\'a}novas}, H. and {Cantat-Gaudin}, T. and {Carballo}, R. and {Carlucci}, T. and {Carnerero}, M.~I. and {Carrasco}, J.~M. and {Casamiquela}, L. and {Castellani}, M. and {Castro-Ginard}, A. and {Castro Sampol}, P. and {Chaoul}, L. and {Charlot}, P. and {Chemin}, L. and {Chiavassa}, A. and {Cioni}, M. -R.~L. and {Comoretto}, G. and {Cooper}, W.~J. and {Cornez}, T. and {Cowell}, S. and {Crifo}, F. and {Crosta}, M. and {Crowley}, C. and {Dafonte}, C. and {Dapergolas}, A. and {David}, M. and {David}, P.},
        title = "{Gaia Early Data Release 3. Summary of the contents and survey properties}",
      journal = {\aap},
         year = 2021,
        month = may,
       volume = {649},
          eid = {A1},
        pages = {A1},
          doi = {10.1051/0004-6361/202039657},
archivePrefix = {arXiv},
       eprint = {2012.01533},
 primaryClass = {astro-ph.GA},
       adsurl = {https://ui.adsabs.harvard.edu/abs/2021A&A...649A...1G}
}

@ARTICLE{2021MNRAS.508.3877G,
       author = {{Gentile Fusillo}, N.~P. and {Tremblay}, P. -E. and {Cukanovaite}, E. and {Vorontseva}, A. and {Lallement}, R. and {Hollands}, M. and {G{\"a}nsicke}, B.~T. and {Burdge}, K.~B. and {McCleery}, J. and {Jordan}, S.},
        title = "{A catalogue of white dwarfs in Gaia EDR3}",
      journal = {\mnras},
         year = 2021,
        month = dec,
       volume = {508},
       number = {3},
        pages = {3877-3896},
          doi = {10.1093/mnras/stab2672},
archivePrefix = {arXiv},
       eprint = {2106.07669},
 primaryClass = {astro-ph.SR},
       adsurl = {https://ui.adsabs.harvard.edu/abs/2021MNRAS.508.3877G}
}

@ARTICLE{2019MNRAS.482.4570G,
       author = {{Gentile Fusillo}, Nicola Pietro and {Tremblay}, Pier-Emmanuel and {G{\"a}nsicke}, Boris T. and {Manser}, Christopher J. and {Cunningham}, Tim and {Cukanovaite}, Elena and {Hollands}, Mark and {Marsh}, Thomas and {Raddi}, Roberto and {Jordan}, Stefan and {Toonen}, Silvia and {Geier}, Stephan and {Barstow}, Martin and {Cummings}, Jeffrey D.},
        title = "{A Gaia Data Release 2 catalogue of white dwarfs and a comparison with SDSS}",
      journal = {\mnras},
         year = 2019,
        month = feb,
       volume = {482},
       number = {4},
        pages = {4570-4591},
          doi = {10.1093/mnras/sty3016},
archivePrefix = {arXiv},
       eprint = {1807.03315},
 primaryClass = {astro-ph.SR},
       adsurl = {https://ui.adsabs.harvard.edu/abs/2019MNRAS.482.4570G}
}

@ARTICLE{2024Natur.627..286B,
       author = {{B{\'e}dard}, Antoine and {Blouin}, Simon and {Cheng}, Sihao},
        title = "{Buoyant crystals halt the cooling of white dwarf stars}",
      journal = {\nat},
         year = 2024,
        month = mar,
       volume = {627},
       number = {8003},
        pages = {286-288},
          doi = {10.1038/s41586-024-07102-y},
archivePrefix = {arXiv},
       eprint = {2409.04419},
 primaryClass = {astro-ph.SR},
       adsurl = {https://ui.adsabs.harvard.edu/abs/2024Natur.627..286B}
}

@ARTICLE{2024A&A...691A.290F,
       author = {{Filiz}, Semih and {Werner}, Klaus and {Rauch}, Thomas and {Reindl}, Nicole},
        title = "{Spectral evolution of hot hybrid white dwarfs: I. Spectral analysis}",
      journal = {\aap},
         year = 2024,
        month = nov,
       volume = {691},
          eid = {A290},
        pages = {A290},
          doi = {10.1051/0004-6361/202451886},
archivePrefix = {arXiv},
       eprint = {2410.14345},
 primaryClass = {astro-ph.SR},
       adsurl = {https://ui.adsabs.harvard.edu/abs/2024A&A...691A.290F}
}

@ARTICLE{2024ApJ...967..166S,
       author = {{Steen}, Maya and {Hermes}, J.~J. and {Guidry}, Joseph A. and {Paiva}, Annabelle and {Farihi}, Jay and {Heintz}, Tyler M. and {Ewing}, Brison B. and {Berry}, Nathaniel},
        title = "{Measuring White Dwarf Variability from Sparsely Sampled Gaia DR3 Multi-epoch Photometry}",
      journal = {\apj},
         year = 2024,
        month = jun,
       volume = {967},
       number = {2},
          eid = {166},
        pages = {166},
          doi = {10.3847/1538-4357/ad3e60},
archivePrefix = {arXiv},
       eprint = {2404.02201},
 primaryClass = {astro-ph.SR},
       adsurl = {https://ui.adsabs.harvard.edu/abs/2024ApJ...967..166S}
}

@ARTICLE{2024ApJ...974..314O,
       author = {{Oliveira da Rosa}, Gabriela and {Kepler}, S.~O. and {Soethe}, L.~T.~T. and {Romero}, Alejandra D. and {Bell}, Keaton J.},
        title = "{Photometric White Dwarf Rotation}",
      journal = {\apj},
         year = 2024,
        month = oct,
       volume = {974},
       number = {2},
          eid = {314},
        pages = {314},
          doi = {10.3847/1538-4357/ad6987},
archivePrefix = {arXiv},
       eprint = {2407.05214},
 primaryClass = {astro-ph.SR},
       adsurl = {https://ui.adsabs.harvard.edu/abs/2024ApJ...974..314O}
}

@ARTICLE{2006PASP..118..183W,
       author = {{Werner}, Klaus and {Herwig}, Falk},
        title = "{The Elemental Abundances in Bare Planetary Nebula Central Stars and the Shell Burning in AGB Stars}",
      journal = {\pasp},
         year = 2006,
        month = feb,
       volume = {118},
       number = {840},
        pages = {183-204},
          doi = {10.1086/500443},
archivePrefix = {arXiv},
       eprint = {astro-ph/0512320},
 primaryClass = {astro-ph},
       adsurl = {https://ui.adsabs.harvard.edu/abs/2006PASP..118..183W}
}

@ARTICLE{2011A&A...536A..43N,
       author = {{Nebot G{\'o}mez-Mor{\'a}n}, A. and {G{\"a}nsicke}, B.~T. and {Schreiber}, M.~R. and {Rebassa-Mansergas}, A. and {Schwope}, A.~D. and {Southworth}, J. and {Aungwerojwit}, A. and {Bothe}, M. and {Davis}, P.~J. and {Kolb}, U. and {M{\"u}ller}, M. and {Papadaki}, C. and {Pyrzas}, S. and {Rabitz}, A. and {Rodr{\'\i}guez-Gil}, P. and {Schmidtobreick}, L. and {Schwarz}, R. and {Tappert}, C. and {Toloza}, O. and {Vogel}, J. and {Zorotovic}, M.},
        title = "{Post common envelope binaries from SDSS. XII. The orbital period distribution}",
      journal = {\aap},
         year = 2011,
        month = dec,
       volume = {536},
          eid = {A43},
        pages = {A43},
          doi = {10.1051/0004-6361/201117514},
archivePrefix = {arXiv},
       eprint = {1109.6662},
 primaryClass = {astro-ph.SR},
       adsurl = {https://ui.adsabs.harvard.edu/abs/2011A&A...536A..43N}
}

@ARTICLE{2007A&A...469..297A,
       author = {{Aungwerojwit}, A. and {G{\"a}nsicke}, B.~T. and {Rodr{\'\i}guez-Gil}, P. and {Hagen}, H. -J. and {Giannakis}, O. and {Papadimitriou}, C. and {Allende Prieto}, C. and {Engels}, D.},
        title = "{HS 1857+5144: a hot and young pre-cataclysmic variable}",
      journal = {\aap},
         year = 2007,
        month = jul,
       volume = {469},
       number = {1},
        pages = {297-305},
          doi = {10.1051/0004-6361:20077276},
archivePrefix = {arXiv},
       eprint = {0704.1780},
 primaryClass = {astro-ph},
       adsurl = {https://ui.adsabs.harvard.edu/abs/2007A&A...469..297A}
}

@ARTICLE{2010MNRAS.402.2591P,
       author = {{Parsons}, S.~G. and {Marsh}, T.~R. and {Copperwheat}, C.~M. and {Dhillon}, V.~S. and {Littlefair}, S.~P. and {G{\"a}nsicke}, B.~T. and {Hickman}, R.},
        title = "{Precise mass and radius values for the white dwarf and low mass M dwarf in the pre-cataclysmic binary NN Serpentis}",
      journal = {\mnras},
         year = 2010,
        month = mar,
       volume = {402},
       number = {4},
        pages = {2591-2608},
          doi = {10.1111/j.1365-2966.2009.16072.x},
archivePrefix = {arXiv},
       eprint = {0909.4307},
 primaryClass = {astro-ph.SR},
       adsurl = {https://ui.adsabs.harvard.edu/abs/2010MNRAS.402.2591P}
}

@ARTICLE{2006MNRAS.365..287B,
       author = {{Brinkworth}, C.~S. and {Marsh}, T.~R. and {Dhillon}, V.~S. and {Knigge}, C.},
        title = "{Detection of a period decrease in NN Ser with ULTRACAM: evidence for strong magnetic braking or an unseen companion}",
      journal = {\mnras},
         year = 2006,
        month = jan,
       volume = {365},
       number = {1},
        pages = {287-295},
          doi = {10.1111/j.1365-2966.2005.09718.x},
archivePrefix = {arXiv},
       eprint = {astro-ph/0510331},
 primaryClass = {astro-ph},
       adsurl = {https://ui.adsabs.harvard.edu/abs/2006MNRAS.365..287B}
}

@ARTICLE{1988ApJ...334..220K,
       author = {{Kawaler}, Steven D.},
        title = "{The Hydrogen Shell Game: Pulsational Instabilities in Hydrogen Shell--Burning Planetary Nebula Nuclei}",
      journal = {\apj},
         year = 1988,
        month = nov,
       volume = {334},
        pages = {220},
          doi = {10.1086/166832},
       adsurl = {https://ui.adsabs.harvard.edu/abs/1988ApJ...334..220K}
}

@INPROCEEDINGS{2017EPJWC.15206012C,
       author = {{Calcaferro}, Leila M. and {C{\'o}rsico}, Alejandro H. and {Camisassa}, Mar{\'\i}a E. and {Althaus}, Leandro G. and {Shibahashi}, Hiromoto},
        title = "{Pulsational instability of high-luminosity H-rich pre-white dwarf star}",
    booktitle = {European Physical Journal Web of Conferences},
         year = 2017,
       series = {European Physical Journal Web of Conferences},
       volume = {152},
        month = sep,
          eid = {06012},
        pages = {06012},
          doi = {10.1051/epjconf/201715206012},
       adsurl = {https://ui.adsabs.harvard.edu/abs/2017EPJWC.15206012C}
}

@ARTICLE{2014PASJ...66...76M,
       author = {{Maeda}, Kazuhiro and {Shibahashi}, Hiromoto},
        title = "{Pulsations of pre-white dwarfs with hydrogen-dominated atmospheres}",
      journal = {\pasj},
         year = 2014,
        month = jul,
       volume = {66},
       number = {4},
          eid = {76},
        pages = {76},
          doi = {10.1093/pasj/psu051},
archivePrefix = {arXiv},
       eprint = {1405.4568},
 primaryClass = {astro-ph.SR},
       adsurl = {https://ui.adsabs.harvard.edu/abs/2014PASJ...66...76M}
}

@ARTICLE{2017ApJ...835..277H,
       author = {{Hermes}, J.~J. and {Kawaler}, Steven D. and {Bischoff-Kim}, A. and {Provencal}, J.~L. and {Dunlap}, B.~H. and {Clemens}, J.~C.},
        title = "{A Deep Test of Radial Differential Rotation in a Helium-atmosphere White Dwarf. I. Discovery of Pulsations in PG 0112+104}",
      journal = {\apj},
         year = 2017,
        month = feb,
       volume = {835},
       number = {2},
          eid = {277},
        pages = {277},
          doi = {10.3847/1538-4357/835/2/277},
archivePrefix = {arXiv},
       eprint = {1612.07807},
 primaryClass = {astro-ph.SR},
       adsurl = {https://ui.adsabs.harvard.edu/abs/2017ApJ...835..277H}
}

@INPROCEEDINGS{2016SPIE.9913E..3EJ,
       author = {{Jenkins}, Jon M. and {Twicken}, Joseph D. and {McCauliff}, Sean and {Campbell}, Jennifer and {Sanderfer}, Dwight and {Lung}, David and {Mansouri-Samani}, Masoud and {Girouard}, Forrest and {Tenenbaum}, Peter and {Klaus}, Todd and {Smith}, Jeffrey C. and {Caldwell}, Douglas A. and {Chacon}, A.~D. and {Henze}, Christopher and {Heiges}, Cory and {Latham}, David W. and {Morgan}, Edward and {Swade}, Daryl and {Rinehart}, Stephen and {Vanderspek}, Roland},
        title = "{The TESS science processing operations center}",
    booktitle = {Software and Cyberinfrastructure for Astronomy IV},
         year = 2016,
       editor = {{Chiozzi}, Gianluca and {Guzman}, Juan C.},
       series = {Society of Photo-Optical Instrumentation Engineers (SPIE) Conference Series},
       volume = {9913},
        month = aug,
          eid = {99133E},
        pages = {99133E},
          doi = {10.1117/12.2233418},
       adsurl = {https://ui.adsabs.harvard.edu/abs/2016SPIE.9913E..3EJ}
}

@ARTICLE{2019PASP..131f8003B,
       author = {{Bellm}, Eric C. and {Kulkarni}, Shrinivas R. and {Barlow}, Tom and {Feindt}, Ulrich and {Graham}, Matthew J. and {Goobar}, Ariel and {Kupfer}, Thomas and {Ngeow}, Chow-Choong and {Nugent}, Peter and {Ofek}, Eran and {Prince}, Thomas A. and {Riddle}, Reed and {Walters}, Richard and {Ye}, Quan-Zhi},
        title = "{The Zwicky Transient Facility: Surveys and Scheduler}",
      journal = {\pasp},
         year = 2019,
        month = jun,
       volume = {131},
       number = {1000},
        pages = {068003},
          doi = {10.1088/1538-3873/ab0c2a},
archivePrefix = {arXiv},
       eprint = {1905.02209},
 primaryClass = {astro-ph.IM},
       adsurl = {https://ui.adsabs.harvard.edu/abs/2019PASP..131f8003B}
}

@ARTICLE{2019PASP..131a8003M,
       author = {{Masci}, Frank J. and {Laher}, Russ R. and {Rusholme}, Ben and {Shupe}, David L. and {Groom}, Steven and {Surace}, Jason and {Jackson}, Edward and {Monkewitz}, Serge and {Beck}, Ron and {Flynn}, David and {Terek}, Scott and {Landry}, Walter and {Hacopians}, Eugean and {Desai}, Vandana and {Howell}, Justin and {Brooke}, Tim and {Imel}, David and {Wachter}, Stefanie and {Ye}, Quan-Zhi and {Lin}, Hsing-Wen and {Cenko}, S. Bradley and {Cunningham}, Virginia and {Rebbapragada}, Umaa and {Bue}, Brian and {Miller}, Adam A. and {Mahabal}, Ashish and {Bellm}, Eric C. and {Patterson}, Maria T. and {Juri{\'c}}, Mario and {Golkhou}, V. Zach and {Ofek}, Eran O. and {Walters}, Richard and {Graham}, Matthew and {Kasliwal}, Mansi M. and {Dekany}, Richard G. and {Kupfer}, Thomas and {Burdge}, Kevin and {Cannella}, Christopher B. and {Barlow}, Tom and {Van Sistine}, Angela and {Giomi}, Matteo and {Fremling}, Christoffer and {Blagorodnova}, Nadejda and {Levitan}, David and {Riddle}, Reed and {Smith}, Roger M. and {Helou}, George and {Prince}, Thomas A. and {Kulkarni}, Shrinivas R.},
        title = "{The Zwicky Transient Facility: Data Processing, Products, and Archive}",
      journal = {\pasp},
         year = 2019,
        month = jan,
       volume = {131},
       number = {995},
        pages = {018003},
          doi = {10.1088/1538-3873/aae8ac},
archivePrefix = {arXiv},
       eprint = {1902.01872},
 primaryClass = {astro-ph.IM},
       adsurl = {https://ui.adsabs.harvard.edu/abs/2019PASP..131a8003M}
}

@ARTICLE{2009ApJ...696..870D,
       author = {{Drake}, A.~J. and {Djorgovski}, S.~G. and {Mahabal}, A. and {Beshore}, E. and {Larson}, S. and {Graham}, M.~J. and {Williams}, R. and {Christensen}, E. and {Catelan}, M. and {Boattini}, A. and {Gibbs}, A. and {Hill}, R. and {Kowalski}, R.},
        title = "{First Results from the Catalina Real-Time Transient Survey}",
      journal = {\apj},
         year = 2009,
        month = may,
       volume = {696},
       number = {1},
        pages = {870-884},
          doi = {10.1088/0004-637X/696/1/870},
archivePrefix = {arXiv},
       eprint = {0809.1394},
 primaryClass = {astro-ph},
       adsurl = {https://ui.adsabs.harvard.edu/abs/2009ApJ...696..870D}
}

@ARTICLE{2018PASP..130f4505T,
       author = {{Tonry}, J.~L. and {Denneau}, L. and {Heinze}, A.~N. and {Stalder}, B. and {Smith}, K.~W. and {Smartt}, S.~J. and {Stubbs}, C.~W. and {Weiland}, H.~J. and {Rest}, A.},
        title = "{ATLAS: A High-cadence All-sky Survey System}",
      journal = {\pasp},
         year = 2018,
        month = jun,
       volume = {130},
       number = {988},
        pages = {064505},
          doi = {10.1088/1538-3873/aabadf},
archivePrefix = {arXiv},
       eprint = {1802.00879},
 primaryClass = {astro-ph.IM},
       adsurl = {https://ui.adsabs.harvard.edu/abs/2018PASP..130f4505T}
}

@ARTICLE{2023A&A...674A..13E,
       author = {{Eyer}, L. and {Audard}, M. and {Holl}, B. and {Rimoldini}, L. and {Carnerero}, M.~I. and {Clementini}, G. and {De Ridder}, J. and {Distefano}, E. and {Evans}, D.~W. and {Gavras}, P. and {Gomel}, R. and {Lebzelter}, T. and {Marton}, G. and {Mowlavi}, N. and {Panahi}, A. and {Ripepi}, V. and {Wyrzykowski}, {\L}. and {Nienartowicz}, K. and {Jevardat de Fombelle}, G. and {Lecoeur-Taibi}, I. and {Rohrbasser}, L. and {Riello}, M. and {Garc{\'\i}a-Lario}, P. and {Lanzafame}, A.~C. and {Mazeh}, T. and {Raiteri}, C.~M. and {Zucker}, S. and {{\'A}brah{\'a}m}, P. and {Aerts}, C. and {Aguado}, J.~J. and {Anderson}, R.~I. and {Bashi}, D. and {Binnenfeld}, A. and {Faigler}, S. and {Garofalo}, A. and {Karbevska}, L. and {K{\'o}sp{\'a}l}, {\'A}. and {Kruszy{\'n}ska}, K. and {Kun}, M. and {Lanza}, A.~F. and {Leccia}, S. and {Marconi}, M. and {Messina}, S. and {Molinaro}, R. and {Moln{\'a}r}, L. and {Muraveva}, T. and {Musella}, I. and {Nagy}, Z. and {Pagano}, I. and {Palaversa}, L. and {Plachy}, E. and {Pr{\v{s}}a}, A. and {Rybicki}, K.~A. and {Shahaf}, S. and {Szabados}, L. and {Szegedi-Elek}, E. and {Trabucchi}, M. and {Barblan}, F. and {Grenon}, M. and {Roelens}, M. and {S{\"u}veges}, M.},
        title = "{Gaia Data Release 3. Summary of the variability processing and analysis}",
      journal = {\aap},
         year = 2023,
        month = jun,
       volume = {674},
          eid = {A13},
        pages = {A13},
          doi = {10.1051/0004-6361/202244242},
archivePrefix = {arXiv},
       eprint = {2206.06416},
 primaryClass = {astro-ph.SR},
       adsurl = {https://ui.adsabs.harvard.edu/abs/2023A&A...674A..13E}
}

@ARTICLE{2020A&A...635A.128A,
       author = {{Aller}, A. and {Lillo-Box}, J. and {Jones}, D. and {Miranda}, L.~F. and {Barcel{\'o} Forteza}, S.},
        title = "{Planetary nebulae seen with TESS: Discovery of new binary central star candidates from Cycle 1}",
      journal = {\aap},
         year = 2020,
        month = mar,
       volume = {635},
          eid = {A128},
        pages = {A128},
          doi = {10.1051/0004-6361/201937118},
archivePrefix = {arXiv},
       eprint = {1911.09991},
 primaryClass = {astro-ph.SR},
       adsurl = {https://ui.adsabs.harvard.edu/abs/2020A&A...635A.128A}
}

@ARTICLE{2023AJ....165..141H,
       author = {{Higgins}, Michael E. and {Bell}, Keaton J.},
        title = "{Localizing Sources of Variability in Crowded TESS Photometry}",
      journal = {\aj},
         year = 2023,
        month = apr,
       volume = {165},
       number = {4},
          eid = {141},
        pages = {141},
          doi = {10.3847/1538-3881/acb20c},
archivePrefix = {arXiv},
       eprint = {2204.06020},
 primaryClass = {astro-ph.IM},
       adsurl = {https://ui.adsabs.harvard.edu/abs/2023AJ....165..141H}
}

@ARTICLE{2024A&A...690A.190A,
       author = {{Aller}, Alba and {Lillo-Box}, Jorge and {Jones}, David},
        title = "{Planetary nebulae seen with TESS: New and revisited short-period binary central star candidates from Cycles 1 to 4}",
      journal = {\aap},
         year = 2024,
        month = oct,
       volume = {690},
          eid = {A190},
        pages = {A190},
          doi = {10.1051/0004-6361/202450942},
archivePrefix = {arXiv},
       eprint = {2409.06332},
 primaryClass = {astro-ph.SR},
       adsurl = {https://ui.adsabs.harvard.edu/abs/2024A&A...690A.190A}
}

@ARTICLE{2024A&A...690A.366R,
       author = {{Reindl}, Nicole and {Bond}, Howard E. and {Werner}, Klaus and {Zeimann}, Gregory R.},
        title = "{Spectroscopic survey of faint planetary-nebula nuclei: VI. Seventeen hydrogen-rich central stars}",
      journal = {\aap},
         year = 2024,
        month = oct,
       volume = {690},
          eid = {A366},
        pages = {A366},
          doi = {10.1051/0004-6361/202451591},
archivePrefix = {arXiv},
       eprint = {2408.01411},
 primaryClass = {astro-ph.SR},
       adsurl = {https://ui.adsabs.harvard.edu/abs/2024A&A...690A.366R}
}

\onecolumn

\begin{appendix} 

\label{Appendix:Tabs}
\section{Tables}

\begin{table*}[h!]
\newcolumntype{e}[1]{D{+}{\,\pm\,}{#1}}
    \caption{Effective temperatures, surface gravities, radii, luminosities, gravity and Kiel masses, color excesses, \emph{Gaia} parallaxes and RUWE factors of DAO and DA WDs.}
    \centering
    \footnotesize
    \begin{tabular}{l e{2.1} c c r c c c e{2.1} r}
        \hline\hline\noalign{\smallskip}
         \multicolumn{1}{>{\centering\arraybackslash}m{1.7cm}}{Name} & \multicolumn{1}{>{\centering\arraybackslash}m{1.7cm}}{\teff} & \multicolumn{1}{>{\centering\arraybackslash}m{1.4cm}}{\logg} & \multicolumn{1}{>{\centering\arraybackslash}m{1.5cm}}{\textit{R}} & \multicolumn{1}{>{\centering\arraybackslash}m{1.3cm}}{\textit{L}} & \multicolumn{1}{>{\centering\arraybackslash}m{1.2cm}}{\Mgrav } & \multicolumn{1}{>{\centering\arraybackslash}m{1.2cm}}{\Mkiel} & \multicolumn{1}{>{\centering\arraybackslash}m{1.7cm}}{\textit{E(44--55)}} & \multicolumn{1}{>{\centering\arraybackslash}m{1.2cm}}{$\varpi_{Gaia}$} & \multicolumn{1}{>{\centering\arraybackslash}m{0.7cm}}{RUWE} \\
          \multicolumn{1}{>{\centering\arraybackslash}m{1.3cm}}{ }& \multicolumn{1}{>{\centering\arraybackslash}m{1.7cm}}{[kK]}& \multicolumn{1}{>{\centering\arraybackslash}m{1.2cm}}{[cm/s$^2$]} &\multicolumn{1}{>{\centering\arraybackslash}m{1.5cm}}{ [\Rsol]} &\multicolumn{1}{>{\centering\arraybackslash}m{1.3cm}}{[\Lsol] } & \multicolumn{1}{>{\centering\arraybackslash}m{1.2cm}}{[\Msol]} & \multicolumn{1}{>{\centering\arraybackslash}m{1.2cm}}{[\Msol]} & \multicolumn{1}{>{\centering\arraybackslash}m{1.7cm}}{[mag]} & \multicolumn{1}{>{\centering\arraybackslash}m{1.2cm}}{[mas]} & \multicolumn{1}{>{\centering\arraybackslash}m{0.7cm}}{ } \\
         \hline\noalign{\smallskip}

         \multicolumn{10}{c}{DAO} \\
Longmore1 	&	118000	+	5000	&	7.0    $\pm$	0.3	&	$0.0419^{+0.0015}_{-0.0014}$	&      $310.0^{+70.0}_{-60.0}$	&	$0.64^{+0.70}_{-0.40}$	&	0.56 $\pm$ 0.06	&	$0.0236^{+0.0024}_{-0.0024}$	&	1.24 + 0.05	&	1.08	\\
\noalign{\smallskip}									       			  																	     
WD\,0439+466	&	97000	+	5000	&	7.0    $\pm$	0.2	&	$0.0243^{+0.0002}_{-0.0002}$	&	$47.0^{+11.0}_{-10.0}$	&	$0.22^{+0.13}_{-0.08}$	&	0.54 $\pm$ 0.06	&	$0.0439^{+0.0028}_{-0.0027}$	&	7.88 + 0.07	&	0.65	\\
\noalign{\smallskip}									       			  																	     
WD\,0500$-$156	&	104000	+	6000	&	7.2    $\pm$	0.2	&	$0.0270^{+0.0007}_{-0.0007}$	&	$77.0^{+20.0}_{-17.0}$	&	$0.42^{+0.25}_{-0.16}$	&	0.57 $\pm$ 0.06	&	$0.0608^{+0.0024}_{-0.0024}$	&	1.95 + 0.05	&	1.06	\\
\noalign{\smallskip}									       			  																	     
WD\,0615+556	&	101000	+	5000	&	7.2    $\pm$	0.2	&	$0.0223^{+0.0007}_{-0.0007}$	&	$47.0^{+11.0}_{-9.0}$	&	$0.29^{+0.18}_{-0.11}$	&	0.57 $\pm$ 0.05	&	$0.1270^{+0.0060}_{-0.0060}$	&	2.55 + 0.07	&	0.98	\\
\noalign{\smallskip}									       			  																	     
WD\,0823+316	&	98000	+	4000	&	7.1    $\pm$	0.2	&	$0.0251^{+0.0009}_{-0.0009}$	&	$53.0^{+13.0}_{-11.0}$	&	$0.29^{+0.18}_{-0.11}$	&	0.55 $\pm$ 0.05	&	$0.0270^{+0.0026}_{-0.0026}$	&	1.81 + 0.07	&	1.02	\\
\noalign{\smallskip}									       			  																	     
WD\,0834+500	&	90000	+	3000	&	7.0    $\pm$	0.3	&	$0.0306^{+0.0010}_{-0.0009}$	&	$55.0^{+9.0}_{-8.0}$	&	$0.34^{+0.21}_{-0.13}$	&	0.53 $\pm$ 0.05	&	$0.0496^{+0.0026}_{-0.0026}$	&	1.95 + 0.06	&	1.03	\\
\noalign{\smallskip}									       			  																	     
WD\,0851+090	&	106000	+	5000	&	7.2    $\pm$	0.2	&	$0.0273^{+0.0010}_{-0.0009}$	&	$85.0^{+21.0}_{-18.0}$	&	$0.43^{+0.28}_{-0.18}$	&	0.57 $\pm$ 0.07	&	$0.0616^{+0.0025}_{-0.0026}$	&	1.86 + 0.06	&	1.03	\\
\noalign{\smallskip}									       			  																	     
WD\,1111+552	&	110000	+	5000	&	7.1    $\pm$	0.3	&	$0.0339^{+0.0015}_{-0.0014}$	&      $152.0^{+33.0}_{-28.0}$	&	$0.53^{+0.54}_{-0.27}$	&	0.56 $\pm$ 0.06	&	$0.0100^{+0.0040}_{-0.0040}$	&	1.23 + 0.05	&	1.04	\\
\noalign{\smallskip}									       			  																	     
WD\,1214+267	&	91000	+	3000	&	7.1    $\pm$	0.3	&	$0.0240^{+0.0009}_{-0.0008}$	&	$35.0^{+6.0}_{-5.0}$	&	$0.26^{+0.27}_{-0.14}$	&	0.54 $\pm$ 0.06	&	$0.0151^{+0.0020}_{-0.0020}$	&	2.10 + 0.08	&	1.27	\\
\noalign{\smallskip}									       			  																	     
WD\,1253+378	&	93000	+	4000	&	7.0    $\pm$	0.2	&	$0.0282^{+0.0012}_{-0.0011}$	&	$54.0^{+13.0}_{-11.0}$	&	$0.29^{+0.19}_{-0.12}$	&	0.53 $\pm$ 0.05	&	$0.0143^{+0.0025}_{-0.0026}$	&	1.61 + 0.06	&	1.20	\\
\noalign{\smallskip}									       			  																	     
WD\,1957+225	&	134000	+	10000	&	6.9    $\pm$	0.4	&	$0.0325^{+0.0006}_{-0.0006}$	&      $310.0^{+110.0}_{-90.0}$	&	$0.31^{+0.47}_{-0.19}$	&	0.56 $\pm$ 0.06	&	$0.0410^{+0.0040}_{-0.0040}$	&	2.57 + 0.05	&	1.16	\\
\noalign{\smallskip}									       			  																	     
WD\,2226$-$210	&	120000	+	5000	&	7.2    $\pm$	0.3	&	$0.0218^{+0.0002}_{-0.0002}$	&	$89.0^{+16.0}_{-14.0}$	&	$0.27^{+0.28}_{-0.14}$	&	0.59 $\pm$ 0.06	&	$0.0102^{+0.0250}_{-0.0250}$	&	5.01 + 0.06	&	0.95	\\
\noalign{\smallskip}									       			  																	     
WD\,2342+806	&	83000	+	5000	&	7.2    $\pm$	0.2	&	$0.0259^{+0.0003}_{-0.0003}$	&	$29.0^{+8.0}_{-7.0}$	&	$0.39^{+0.23}_{-0.15}$	&	0.53 $\pm$ 0.05	&	$0.0638^{+0.0022}_{-0.0022}$	&	3.45 + 0.04	&	1.01	\\

\noalign{\smallskip}
    \hline\noalign{\smallskip}
    \multicolumn{10}{c}{DA} \\
WD\,0027$-$636	&	59000	+	5000	&	8.0    $\pm$	0.2	&	$0.0146^{+0.0001}_{-0.0001}$	&	 $2.3^{+0.9}_{-0.7}$	&	$0.77^{+0.46}_{-0.29}$	&	0.69 $\pm$ 0.07	&	$0.0137^{+0.0025}_{-0.0025}$	&	5.09 + 0.04	&	1.01	\\
\noalign{\smallskip}									       			  																	     
WD\,0229$-$481	&	62000	+	5000	&	7.8    $\pm$	0.2	&	$0.0171^{+0.0002}_{-0.0001}$	&	 $3.9^{+1.5}_{-1.2}$	&	$0.67^{+0.40}_{-0.25}$	&	0.62 $\pm$ 0.07	&	$0.0100^{+0.0040}_{-0.0040}$	&	5.93 + 0.04	&	1.04	\\
\noalign{\smallskip}									       			  																	     
WD\,0232+035	&	63000	+	3000	&	7.5    $\pm$	0.2	&	$0.0207^{+0.0003}_{-0.0003}$	&	 $6.1^{+1.4}_{-1.2}$	&	$0.49^{+0.32}_{-0.20}$	&	0.54 $\pm$ 0.07	&	$0.0390^{+0.0040}_{-0.0040}$	&      12.91 + 0.05	&	1.58	\\
\noalign{\smallskip}									       			  																	     
WD\,0311+480	&	70000	+	5000	&	7.3    $\pm$	0.4	&	$0.0285^{+0.0004}_{-0.0004}$	&	$18.0^{+13.0}_{-9.0}$	&	$0.60^{+0.90}_{-0.40}$	&	0.52 $\pm$ 0.11	&	$0.1360^{+0.0050}_{-0.0050}$	&	4.34 + 0.05	&	1.14	\\
\noalign{\smallskip}									       			  																	     
WD\,0343$-$007	&	63000	+	4000	&	7.7    $\pm$	0.2	&	$0.0183^{+0.0002}_{-0.0002}$	&	 $4.7^{+1.4}_{-1.1}$	&	$0.61^{+0.36}_{-0.23}$	&	0.59 $\pm$ 0.07	&	$0.0547^{+0.0017}_{-0.0017}$	&	4.96 + 0.06	&	1.08	\\
\noalign{\smallskip}									       			  																	     
WD\,0455$-$282	&	66000	+	3000	&	7.5    $\pm$	0.2	&	$0.0166^{+0.0007}_{-0.0007}$	&	 $4.7^{+1.1}_{-0.9}$	&	$0.32^{+0.20}_{-0.12}$	&	0.55 $\pm$ 0.05	&	$0.0180^{+0.0040}_{-0.0040}$	&	8.00 + 0.40	&	8.65	\\
\noalign{\smallskip}									       			  																	     
WD\,0615+655	&	83000	+	10000	&	7.6    $\pm$	0.4	&	$0.0171^{+0.0003}_{-0.0003}$	&	$13.0^{+8.0}_{-6.0}$	&	$0.43^{+0.65}_{-0.26}$	&	0.61 $\pm$ 0.10	&	$0.0930^{+0.0040}_{-0.0040}$	&	2.92 + 0.05	&	0.97	\\
\noalign{\smallskip}									       			  																	     
WD\,0621$-$376	&	65000	+	3000	&	7.5    $\pm$	0.4	&	$0.0232^{+0.0001}_{-0.0001}$	&	 $8.7^{+1.8}_{-1.5}$	&	$0.60^{+1.00}_{-0.40}$	&	0.54 $\pm$ 0.10	&	$0.0051^{+0.0018}_{-0.0018}$	&      13.09 + 0.05	&	0.97	\\
\noalign{\smallskip}									       			  																	     
WD\,0939+262	&	66000	+	3000	&	7.7    $\pm$	0.2	&	$0.0173^{+0.0002}_{-0.0002}$	&	 $5.1^{+1.0}_{-0.9}$	&	$0.54^{+0.32}_{-0.21}$	&	0.60 $\pm$ 0.06	&	$0.0298^{+0.0023}_{-0.0023}$	&	5.46 + 0.06	&	1.14	\\
\noalign{\smallskip}									       			  																	     
WD\,1056+516	&	63000	+	5000	&	7.9    $\pm$	0.2	&	$0.0110^{+0.0003}_{-0.0002}$	&	 $1.7^{+0.7}_{-0.5}$	&	$0.35^{+0.21}_{-0.13}$	&	0.66 $\pm$ 0.07	&	$0.0127^{+0.0023}_{-0.0023}$	&	3.41 + 0.08	&	0.97	\\
\noalign{\smallskip}									       			  																	     
WD\,1342+443	&	62000	+	5000	&	7.7    $\pm$	0.3	&	$0.0167^{+0.0005}_{-0.0005}$	&	 $3.7^{+1.4}_{-1.1}$	&	$0.51^{+0.52}_{-0.26}$	&	0.59 $\pm$ 0.08	&	$0.0183^{+0.0015}_{-0.0015}$	&	2.33 + 0.07	&	1.08	\\
\noalign{\smallskip}									       			  																	     
WD\,1738+669	&	78000	+	6000	&	7.6    $\pm$	0.4	&	$0.0154^{+0.0001}_{-0.0001}$	&	 $7.9^{+2.8}_{-2.2}$	&	$0.34^{+0.52}_{-0.21}$	&	0.60 $\pm$ 0.10	&	$0.0156^{+0.0018}_{-0.0018}$	&	5.82 + 0.04	&	1.07	\\
\noalign{\smallskip}									       			  																	     
WD\,1827+778	&	78000	+	10000	&	7.4    $\pm$	0.4	&	$0.0212^{+0.0004}_{-0.0004}$	&	$15.0^{+10.0}_{-7.0}$	&	$0.41^{+0.63}_{-0.25}$	&	0.55 $\pm$ 0.09	&	$0.0989^{+0.0030}_{-0.0030}$	&	2.31 + 0.04	&	1.01	\\
\noalign{\smallskip}									       			  																	     
WD\,2046+396	&	64000	+	5000	&	7.8    $\pm$	0.3	&	$0.0164^{+0.0001}_{-0.0001}$	&	 $4.0^{+1.5}_{-1.2}$	&	$0.60^{+0.70}_{-0.40}$	&	0.63 $\pm$ 0.09	&	$0.0244^{+0.0027}_{-0.0027}$	&	6.76 + 0.04	&	1.10	\\
\noalign{\smallskip}									       			  																	     
WD\,2146$-$433	&	66000	+	4000	&	7.5    $\pm$	0.3	&	$0.0197^{+0.0004}_{-0.0004}$	&	 $6.7^{+1.8}_{-1.5}$	&	$0.45^{+0.45}_{-0.23}$	&	0.55 $\pm$ 0.08	&	$0.0208^{+0.0029}_{-0.0029}$	&	2.93 + 0.05	&	1.00	\\
\noalign{\smallskip}									       			  																	     
WD\,2211$-$495	&	68000	+	4000	&	7.4    $\pm$	0.3	&	$0.0211^{+0.0001}_{-0.0000}$	&	 $8.6^{+2.3}_{-1.9}$	&	$0.41^{+0.41}_{-0.21}$	&	0.53 $\pm$ 0.08	&	$0.0005^{+0.0018}_{-0.0005}$	&      17.02 + 0.06	&	1.28	\\
\noalign{\smallskip}									       			  																	     
WD\,2218+706	&	78000	+	5000	&	7.4    $\pm$	0.3	&	$0.0249^{+0.0010}_{-0.0008}$	&	$20.1^{+7.0}_{-6.0}$	&	$0.57^{+0.70}_{-0.40}$	&	0.55 $\pm$ 0.07	&	$0.1640^{+0.0140}_{-0.0120}$	&	2.99 + 0.04	&	1.22	\\
\noalign{\smallskip}									       			  																	     
WD\,2350$-$706	&	75000	+	5000	&	7.9    $\pm$	0.3	&	$0.0155^{+0.0008}_{-0.0004}$	&	 $6.9^{+2.6}_{-2.1}$	&	$0.70^{+0.90}_{-0.50}$	&	0.68 $\pm$ 0.09	&	$0.0020^{+0.0088}_{-0.0020}$	&	6.69 + 0.03	&	1.57	\\
\noalign{\smallskip}									       			  																	     
WD\,2353+026	&	61000	+	5000	&	7.5    $\pm$	0.3	&	$0.0164^{+0.0004}_{-0.0004}$	&	 $3.4^{+1.3}_{-1.0}$	&	$0.31^{+0.32}_{-0.16}$	&	0.53 $\pm$ 0.08	&	$0.0325^{+0.0028}_{-0.0028}$	&	3.64 + 0.08	&	1.11	\\
\noalign{\smallskip}
\hline\
     \end{tabular}
     \tablefoot{\teff, \logg, and \Mkiel were derived in Paper I, whereas \textit{R}, \textit{L}, and \Mgrav were derived in this paper.}
    \label{tab:RML}
\end{table*}

\newpage

\label{Appendix:Figs}
\section{Figures}

\begin{figure*}[h!]
 \centering
 \includegraphics[width=17.5cm]{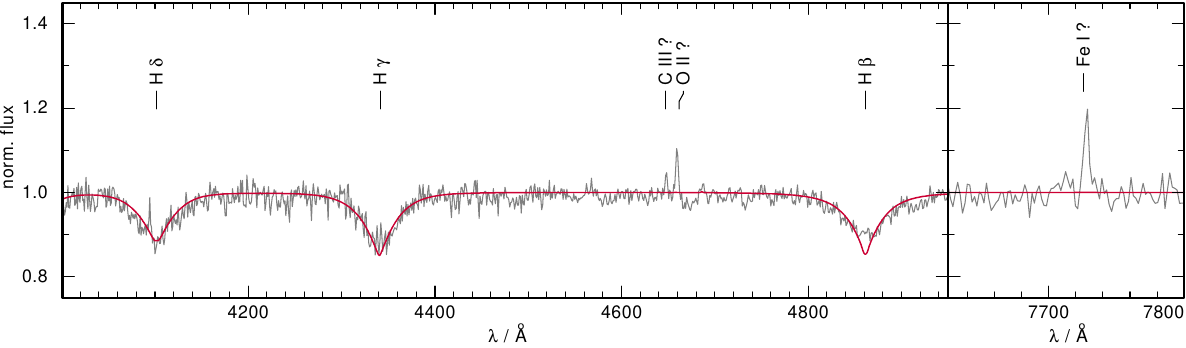}
   \caption{BOSS spectrum of WD\,1342+443 (grey) and the best fit TMAP model (red, \teff = 62 kK and \logg = 7.7) from Paper I.}
     \label{fig:BOSS_wd1342}
 \end{figure*}

 \begin{figure*}[h!]
 \centering
 \includegraphics[width=17.5cm]{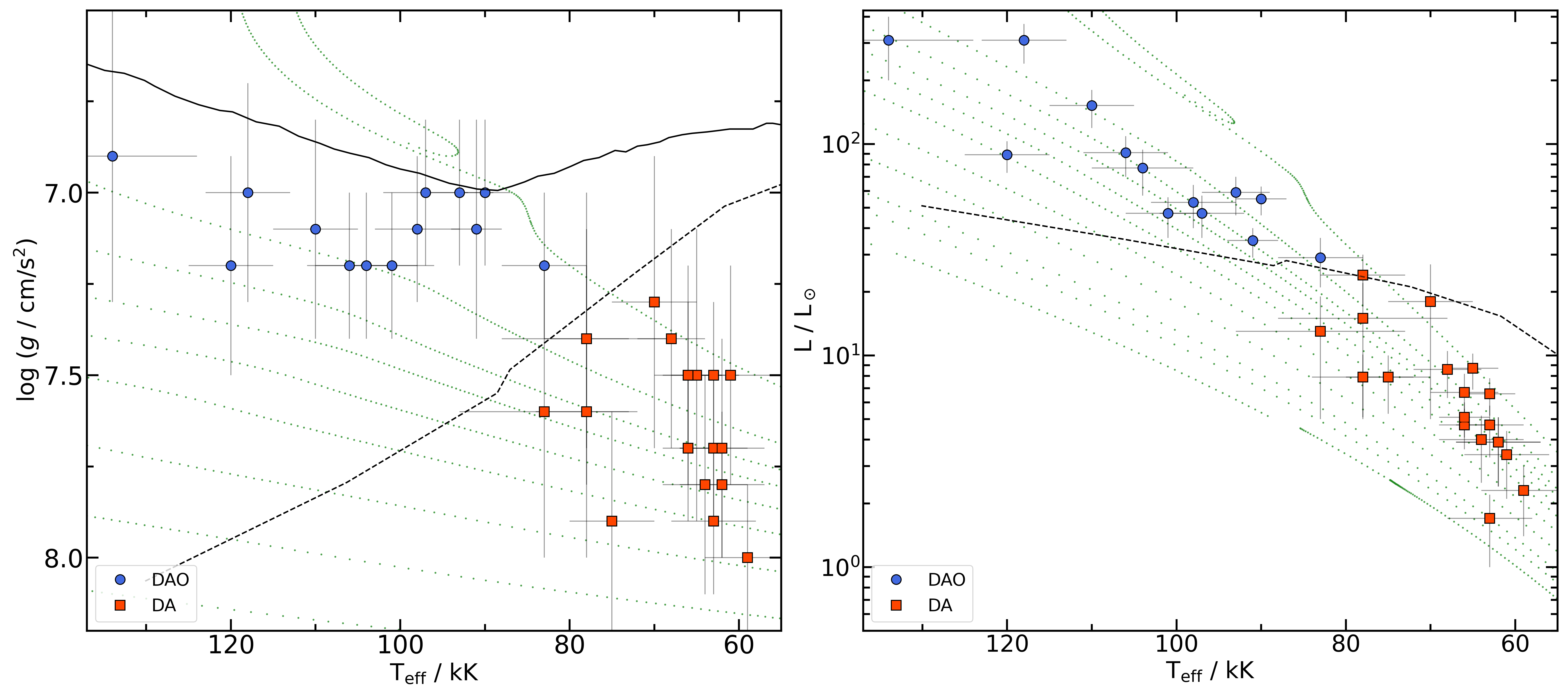}
   \caption{Sample WDs shown in the Kiel diagram (left) and HRD (right). Solid black line in the left panel corresponds to the theoretical wind limit by \citep{2000A&A...359.1042U}. Dashed black lines in both panels represent predicted He abundance (\mbox{\textit{N}(He)/\textit{N}(H) = 10$\textsuperscript{-3}$}) in the cooling sequence by the same authors, which also matches with the approximate optical detection limit of He. Green dotted lines (left: 0.525, 0.570, 0.593, 0.609, 0.632, 0.659, 0.705, 0.767, and 0.837 \Msol, right: 0.525 - 0.935 \Msol) represent evolutionary tracks of H-rich WDs (Z = 0.01) from \citet{2010ApJ...717..183R}. }
     \label{fig:Kiel_HRD}
 \end{figure*}

\end{appendix}

\end{document}